# Sequential Voronoi diagram calculations using simple chemical reactions

B. P. J de Lacy Costello*, I. Jahan and A. Adamatzky

Unconventional computing Group, University of the West of England, Bristol, UK.

*Ben.DeLacyCostello@uwe.ac.uk

In our recent paper [de Lacy Costello *et al*. 2010] we described the formation of complex tessellations of the plane arising from the various reactions of metal salts with potassium ferricyanide and ferrocyanide loaded gels. In addition to producing colourful tessellations these reactions are naturally computing generalised Voronoi diagrams of the plane. The reactions reported previously were capable of the calculation of three distinct Voronoi diagrams of the plane. As diffusion coupled with a chemical reaction is responsible for the calculation then this is achieved in parallel. Thus an increase in the complexity of the data input does not utilise additional computational resource. Additional benefits of these chemical reactions is that a permanent record of the Voronoi diagram calculation (in the form of precipitate free bisectors) is achieved, so there is no requirement for further processing to extract the calculation results. Previously it was assumed that the permanence of the results was also a potential drawback which limited reusability. This paper presents new data which shows that sequential Voronoi diagram calculations can be performed on the same chemical substrate. This is dependent on the reactivity of the original reagent and the cross reactivity of the secondary reagent with the primary product. We present the results from a number of binary combinations of metal salts on both potassium ferricyanide and potassium ferrocyanide substrates. We observe three distinct mechanisms whereby secondary sequential Voronoi diagrams can be calculated. In most cases the result was two interpenetrating permanent Voronoi diagrams. This is interesting from the perspective of mapping the capability of unconventional computing substrates. But also in the study of natural pattern formation *per se*.

## 1. Introduction

Voronoi diagrams are a partitioning of continuous space into regions. The regions are constructed about a finite set of distinct isolated points. Therefore, all locations within the space are associated with the closest member of the point set [Okabe et al. 2009]. Thus, Voronoi diagrams have utility as modelling tools where space should be segregated into "spheres of influence." They find uses in many fields such as astronomy (galaxy cluster analysis [Pastzor & Csillag,1994]), biology (modeling tumour cell growth [Blackburn & Dunckley, 1996]), chemistry (modelling crystal growth [Hargittai, 1986], ecology (modelling competition [Byers, 1992] and computational geometry (solving nearest neighbor problems [Graham & Yao, 1990]. In addition to standard and generalized Voronoi diagrams (where geometric shapes rather than point sources are utilized) weighted diagrams are commonly used, for example, to model crystal growth [Scaudt & Drysdale, 1991].

In addition to their many uses in modeling, Voronoi diagrams can be observed widely in nature, for example, in animal coat markings [Walter *et al.*, 1998] between interacting bacterial, fungal or slime mould colonies [Aurenhammer, 1991; Muller *et al.*, 1998, Shirikawa *et al.* 2009], between growing crystals [de Lacy Costello et al. 2010] and even in universal structures such as gravitational caustics [Stevens, 1979]. They have also been observed and constructed within Gas discharge systems [Zanin *et al.*, 2002 and de Lacy Costello *et al.* 2004].

The formation of Voronoi diagrams in chemical systems with one reagent and one substrate was first reported by [Tolmachiev & Adamatzky, 1996]. This and similar reactions were utilized as chemical processors for the computation of the shortest obstacle free path [Adamatzky & de Lacy Costello, 2003], the skeletonization of a planar shape [Adamatzky & Tolmachiev, 1997; Adamatzky *et al.*, 2002] and the construction of a prototype XOR gate [Adamatzky & de Lacy Costello, 2002]. The formation of weighted Voronoi diagrams from two reagents reacted in parallel on one substrate were first reported in [de Lacy Costello, 2003]. In the same year the calculation of three Voronoi diagrams in parallel was reported in a chemical system based on the reaction of two reagents on a mixed substrate gel (potassium ferrocyanide and potassium ferricyanide) [de Lacy Costello & Adamatzky, 2003]. Complex tessellations of the plane have been constructed in simple chemical reactions where one of two binary reagents exhibited limited or no reactivity with the gel substrate [de Lacy Costello *et al.*,2009]. A more recent paper investigated complex tessellations arising from three reagents on certain substrate loaded gels [ de Lacy Costello *et al.* 2010].

A beneficial feature of the chemical reactions used to create Voronoi diagrams is that when the reaction is completed a "permanent" record of the calculation remains in the form of precipitate free bisectors surrounding coloured cells. Other algorithms for Voronoi diagram calculation require additional processing to extract the final result. It was assumed that these "chemical processors" were single use. Therefore, despite complex calculations being accomplished in parallel of upto three individual Voronoi diagrams it was assumed that additional calculations were not possible. This paper explores binary combinations of metal salts and their reaction with potassium ferrocyanide and potassium ferricyanide loaded gels. We are able to show that for certain combinations of reagents the reactions are reuseable and capable of sequential Voronoi diagram calculations. Voronoi diagrams could be constructed by regressing edges of a spontaneous precipitation/crystallisation reaction but this reaction was not reusable due to the use of a gel substrate[de Lacy Costello *et al*. 2004b] . More recently it was shown that re-useable processors for calculation of Voronoi diagrams could be constructed using a crystallisation process [Adamatzky 2009]. However, these processors must be reset completely prior to carrying out sequential calculations. Sequential calculations are useful in comparing overlapping Voronoi regions. The outputs of this class of reaction are also interesting in terms of a fundamental study of pattern forming mechanisms in natural systems.

In this paper we present new results which show that it is possible to generate sequential Voronoi diagrams on a single substrate provided that appropriate reagents are selected. The paper presents results for all binary reactions of nine distinct metal ions reacted on two distinct chemical substrates (potassium ferrocyanide and potassium ferricyanide loaded gels).

## 2. Experimental

### 2.1. *Formation of sequential Voronoi diagrams on potassium ferrocyanide and potassium ferricyanide gels*

Agar Gel (Sigma, St Louis, USA, 0.3 g) was added to deionized water (30 ml) and heated with stirring to 70◦C and then removed from the heat. Potassium ferricyanide (Fisher Scientific, Leicestershire, UK, 75mg (7.59mM)) or Potassium ferrocyanide (BDH chemicals Ltd, Poole, UK, 75mg (5.91mM)) was added with stirring. The solution was then poured into five 9 cm diameter Petri dishes and left to set for one hour. All possible binary combinations of the metal ions listed below were reacted on both potassium ferrocyanide and ferricyanide gels. The metal ions used in construction of the tessellations were as follows: Iron (III) nitrate [300mg/ml, 1.24 M] (can be substituted with chloride salt), Iron (II) sulphate [300mg/ml, 1.97 M], silver (III) nitrate [300mg/ml, 1.76 M], cobalt (II) chloride hexahydrate [300mg/ml, 1.26 M], lead (II) chloride [300mg/ml, 1.07 M], manganese (II) chloride tetrahydrate [300mg/ml, 1.51 M], chromium (III) nitrate nonahydrate [300mg/ml, 1.26 M], nickel (II) chloride [300mg/ml, 2.31 M] and copper(II) chloride [300mg/ml, 2.23 M].

Voronoi diagrams were formed by placing drops of each metal ion listed above in irregular patterns onto the surface of the potassium ferrocyanide (or ferricyanide) loaded substrate gels. Once the initial reaction was complete a second set of drops of one of the remaining metal ions were placed on the previously reacted gel surface in a different irregular pattern. This process was repeated until each metal ion listed above had been reacted as the primary and secondary reactant on each of the substrate gels. This enabled a classification of the number of Voronoi diagrams calculated with each set of binary reactants. If the first reactant formed a Voronoi diagram but a secondary diagram was not formed by the reagent a [1,0] classification was attributed. However, if the 1$^{st}$ reactant failed to produce a Voronoi diagram but the 2$^{nd}$ reagent did establish a completed Voronoi diagram then a [0,1] classification was attributed. If both sets of reagents produced Voronoi diagrams then a [1,1] classification was attributed. However, if both reactants were unreactive with the substrate then this was given a [0,0] classification. If only partial primary or secondary Voronoi diagrams were formed then this was given a [-,1] or [1,-] classification.

The final reactions were monitored to assess the formation of any secondary Voronoi diagrams. Scans of selected reactions were taken using a flatbed scanner (HP Scanjet 5590) attached to a PC. Scans of certain reactions whilst in progress were recorded at regular intervals in order that the kinetics of the reaction could be observed.

## 3. Results and Discussion

**Tables 1** and **2** detail the Voronoi diagrams formed from all possible binary combinations of the selected metal ions reacted on potassium ferrocyanide and potassium ferricyanide gels respectively. If the primary reactants are the same as the secondary reactants then the result is always [1,0] if the metal ions are reactive with the gel susbtrate (cells highlighted red in **Table 1** and **2**). If the metal ions are unreactive the result is always [0,0] for example where chromium or lead ions are used as the primary and secondary reactants (cells highlighted blue in **Table 1** and **2**). If binary combinations of reagents are considered then the most common result is [1,0] (cells not highlighted), this means that most primary reactants are reactive with the gel (forming Voronoi diagrams) and the majority of secondary reactants are unreactive with respect to the products of the original reaction. If the primary reactant is unreactive but the secondary reagent is reactive then the output [0,1] is obtained. The result is qualitatively similar in that one Voronoi diagram is formed, but the mechanism of formation is distinct.

If potassium ferrocyanide gel is considered then only six combinations give an output [1,1] meaning that two distinct and complete Voronoi diagrams are formed from each set of metal ions/ points (highlighted green in **Table 1**). A further ten combinations give the output [1,-] indicating an incomplete formation of the second Voronoi diagram. Generally outputs [1,1] and [1,-] are most common where silver ions are either the primary (three incidences of [1,1] and one incidence of [1,0]) or secondary reactant (two incidences of [1,1] and three incidences of [1,-]. Where $Fe^{2+}$ or $Fe^{3+}$ ions are the secondary reactant then [1,-] is a common output. This is laos true of reactions on potassium ferricyanide gel. This is probably because both $Fe^{3+}$ and $Fe^{2+}$ ions are known to have strong reactions with potassium ferrocyanide and potassium ferricyanide gel forming Prussian blue and Turnbulls blue respectively. Therefore, metal ions with a weaker interaction with the gel substrate are likely to be displaced in subsequent reactions with $Fe^{3+}$ and $Fe^{2+}$. These reactions invariably give a secondary Voronoi diagram with no permanent output as shown in **Figure 1**.

If potassium ferricyanide gel is considered then the output [1,1] is more prevalent with nine incidences (cells marked in green in **Table 2**). There are also a further fourteen incidences of the output [1,-] (cells marked in yellow in **Table 2**). In common with the potassium ferrocyanide results a trend of double Voronoi diagram formation exists where silver ions are either the primary (four incidences of [1,1] and two of [1,-] or secondary (one incidence of [1,1] and three of [1,-]) reactant. However, in contrast to the potassium ferrocyanide results a number of [1,1] outputs exists where $Fe^{3+}$ ions are the primary reactant. The reason for this is that ferric ions have a weak reaction with potassium ferricyanide gel. Therefore, despite forming a Voronoi diagram the primary product is cross reactive with a number of the other bivalent ions ($Co^{2+}$, $Ni^{2+}$, $Cu^{2+}$ and $Fe^{2+}$) allowing the formation of a secondary Voronoi diagram. However, the opposite should be true i.e. ferrous ions should be less reactive with potassium ferrocyanide gel. Therefore, we might expect the formation of double Voronoi diagrams where ferrous ions are the primary reactant on potassium ferrocyanide gels. But this is not found experimentally suggesting a differential relative strength of the interactions. The interactions of $Fe^{3+}$ and $Ag^{+}$ do highlight the expected differential behaviour when considering the two substrate gels. In **Table 1**. Where $Fe^{3+}$ is the primary reactant and $Ag^{+}$ is the secondary reactant this is the only incidence of the output [1,1]. However, in **Table 2** where $Fe^{3+}$ is the primary reactant and $Ag^{+}$ is the secondary reactant an output of [1,0] is obtained. However, despite the majority of the behaviours being distinct when comparing the reactions on the two substrates, the reaction of $Ag^{+}$ (primary reactant) and $Co^{2+}$ and $Cu^{2+}$ (secondary reactants) are the same regardless of substrate.

It is clear from the results that the order of reaction is important to the outcome, therefore, the results are rarely interchangeable if the primary and secondary ions are reversed. Take for example the reaction of $Ag^{+}$ and $Mn^{2+}$ and $Co^{2+}$ on potassium ferrocyanide gel (Table 1), the output is [1,1] where $Ag^{+}$ is the primary reactant but [1,-] when it is the secondary reactant. The same is true for the reaction of $Co^{2+}$ and $Ni^{2+}$ on potassium ferricyanide gel.

What is also clear from the results is a fairly predictable pattern of outputs is established on both substrates enabling appropriate reagents to be selected in experimental design. Therefore, it is possible to select a number of reagent pairs which are capable of reproducible construction of two Voronoi diagrams. However, it would have been difficult to predict all aspects of the behaviour without running the binary classification experiments which we have reported. In loose terms it is based on chemical reactivity and common substitution reactions but that said many reactions of metal ions with ferrocyanide and ferricyanide substrates are complex and have not been extensively studied. This is particularly true of binary reactions of metal ions on the same substrate.

The 6 reagent pairs identified in **Table 1** and 9 in **Table 2**. constitute a new class of pattern forming reaction capable of the sequential calculation of two distinct overlapping Voronoi diagrams. In the past this type of tessellation had only been produced in parallel via the careful selection of reagents and positioning of the original sites. The reagent sets identified here allow a programmable pattern formation. They constitute another milestone in the exploration of the computational abilities of unconventional computing substrates. This work has only investigated binary combinations of reagents and it is difficult to predict whether tertiary or higher order combinations of reagents could be selected to accomplish more complex overlapping tessellations of the plane. There is no barrier to the use of these reactions in 3 dimensions apart from a problem of experimental design and a difficulty in data extraction, however, the possibility remains. A beauty of these chemical systems is that within the same reaction area thousands of data inputs can be added without any increase in computational resource or overall computation time. In fact often computational time is reduced as seemingly complexity is increased. The examples we have given show widely spaced drops placed strategically on gel substrates in order to highlight the pattern forming mechanism and to aid visual assessment of the output. However, the limit of the reactions is yet to be established although reaction of nanolitre sized drops has been shown to successfully create Voronoi diagrams.

These reactions can be considered as precipitation reactions generally. However, they are obviously distinct from crystallisation reactions which have shown utility in the construction of Voronoi diagrams and certain weighted Voronoi diagrams. Actually the results shown here highlight this subtle difference. In a crystallisation reaction a state change is induced at a number of given points and the interaction of the growing fronts results in the construction of a Voronoi diagram. If they are all initiated at the same time and grow at the same rate then the Voronoi diagram is a standard generalised type. If they are initiated at distinct times or grow at distinct rates then an additively weighted or Multiplicatively weighted Voronoi diagram is constructed. To reverse the reaction and reuse it is simply a case of reversing the state change via dissolution or heating etc. However, inputting further data once the reaction is complete would seem almost impossible given the physical nature of the reaction. The more subtle chemical mechanism of formation presented here allows for sequential Voronoi diagram calculation. This is because products are formed which are selectively cross-reactive with additional reagents. We actually observe three distinct mechanisms by which two distinct Voronoi diagrams can be said to have formed. However, only two of the mechanisms lead to the formation of two permanent Voronoi diagrams (which is the focus of the current paper). The other mechanism is simply a colour change or bleaching mechanism. Thus a second reagent added to the first completed Voronoi diagram causes a change in the colour of the original precipitate as it diffuses from the reaction site. An example of this is shown for the reaction of $Cu^{2+}$ with potassium ferricyanide gel and subsequently the reaction of $Fe^{2+}$ with the completed primary Voronoi diagram (**Figure 1a-c**) However, where diffusing fronts meet no "permanent" Voronoi diagram is formed. However, in certain algorithms for the construction of a Voronoi diagram a permanent output is not naturally produced but constructed from the intersection points of the interacting fronts. For example the case of Voronoi diagram construction using interacting excitation fronts in the BZ reaction [Adamatzky and de Lacy Costello 2002b]. Therefore, as with other methods a secondary Voronoi diagram could be extracted from the results of these reactions with additional processing. Also if we ignore a computational aim these reactions constitute an interesting class of pattern forming reactions worthy of further study.

There are two additional main mechanisms of sequential Voronoi diagram formation. The first mechanism is where the primary reagent has a limited reactivity with the gel. Despite this limited reactivity it constructs a Voronoi diagram of the plane as it diffuses from the sites and the opposing fronts interact. However, reagents with a stronger reaction with the gel substrate are able to form a second Voronoi diagram when placed sequentially on the gel substrate. The reactions where $Fe^{3+}$ ions are the primary reagent source and potassium ferricyanide is the substrate constitute the main examples of this mechanism (see **Figure 2a-e**).

The second mechanism involves predominantly the reaction of silver ions with both substrates. Most products of the chemical reactions excluding silver are brightly coloured metal ferrocyanides (or ferricyanides) which are translucent. This also follows for silver ions when reacted with potassium ferrocyanide (brown/purple product) and potassium ferricyanide (orange product) substrates. However, when seconday ions are added to reactions involving silver ions a physical precipitate is formed as the reagents diffuse and react. An example of the type of Voronoi structures formed and their mechanism of formation where $Ag^{+}$ are the primary reactants is shown in **Figure 3 a-e.** Examples of other sequential Voronoi diagram formation where Ag+ ions are the primary reactant are shown in **Figure 4a-f**. The same type of mechanism is observed when silver is the secondary ion, however, often diffusion is limited leading to incomplete tessellation formation (see **Figure 5a**). Other examples of the incomplete formation of secondary Voronoi diagrams are shown in **Figure 5b-d**. Reactions which do form secondary Voronoi diagrams where $Ag^{+}$ are the secondary reactant include the reaction of $Ni^{2+}$ and $Ag^{+}$ on potassium ferrocyanide gel (**Figure 6a**). $Fe^{3+}$ reacted with $Ag^{+}$ on potassium ferrocyanide gel (**Figure 6b**) and $Fe^{2+}$ reacted with $Ag^{+}$ on potassium ferricyanide gel also form two complete Voronoi diagrams. Apart from the physical nature of the precipitation it can be seen how the mechanisms differ, as sequential Voronoi diagrams are formed in these systems despite a strong initial reaction between the primary reagent and the gel substrate. Two main mechanisms for bisector formation in chemical systems have been proposed, one based on substrate competition between the advancing fronts, one based on a supercritical threshold at the boundary causing redissolution of the formed product. The physical nature of the precipitation in the silver based reactions especially coupled with the lack of any noticeable redissolution of the formed product at the point of reactant addition leads to the conclusion that the major mechanism of Voronoi diagram (bisector) construction observed in these systems is via substrate competition.

The use of two interacting Voronoi diagrams has found utility in modelling the influence of fluorescent probes on the structure of lipid bilayers [Repakova *et al.* 2005]. Voronoi diagrams constructed with overlapping regions have also shown utility in modelling facility locations [Drezner and Drezner 2012]. Therefore, there are many modelling applications which require an analysis of interacting Voronoi diagrams. Although the chemical systems reported here are capable of sequential construction of interacting Voronoi diagrams they would not have any utility in modelling directly in their current form. However, a study of the formation in natural systems can inform future models in order to make them more robust. A good example of this is the multiplicatively Weighted Crystal Growth Voronoi Diagram which is an extension of the multiplicatively weighted derivative that overcomes the problem of disconnected regions which are not possible in growing crystals [Scaudt and Drysdale, 1991] The greatest significance of these structures remains as a new class of pattern forming reactions. It is possible that similar overlapping Voronoi structures may be observed in nature, although not necessarily in this exact form. Thus a study of the patterns and their mechanism of formation may help in establishing mechanisms in natural systems.

## Conclusions

We report the programmable construction of two sequential Voronoi diagrams in a simple chemical system. An extensive study of the binary sequential reactions of nine separate metal ions on two distinct substrates highlighted a number of pairings capable of constructing two permanent overlapping Voronoi diagrams.
These chemical processors are capable of these complex geometric calculations in parallel. Increasing the complexity of the problem (3D, non standard initiation sites, more data points in a defined area, 2+ reagents) does not increase the computational requirements or computing time. The comparison of overlapping Voronoi regions is a useful computational tool.
In addition to their usefulness in modelling Voronoi diagram formation in natural systems, this set of reactions constitute an interesting class of pattern forming reactions in their own right. The study of these reactions may help to determine the underlying mechanisms of natural pattern formation especially as Voronoi type tessellations are relatively common in nature.

## Acknowledgements

The authors wish to acknowledge funding for Ishrat Jahan from the EPSRC (grant number EP/E016839/1).

| Metal ion | | Primary reactant ions | | | | | | | | |
|---|---|---|---|---|---|---|---|---|---|---|
| | | $Mn^{2+}$ | $Co^{2+}$ | $Ag^{+}$ | $Ni^{2+}$ | $Cu^{2+}$ | $Fe^{2+}$ | $Fe^{3+}$ | $Pb^{2+}$ | $Cr^{3+}$ |
| Secondary reactant ions | $Mn^{2+}$ | 1,0 | 1,0 | 1,1 | 1,0 | 1,0 | 1,0 | 1,0 | 0,1 | 0,1 |
| | $Co^{2+}$ | 1,0 | 1,0 | 1,1 | 1,0 | 1,0 | 1,0 | 1,0 | 0,1 | 0,1 |
| | $Ag^{+}$ | 1, - | 1, - | 1,0 | 1,1 | 1,0 | 1, - | 1,1 | 0,1 | 0,1 |
| | $Ni^{2+}$ | 1,0 | 1,0 | 1,0 | 1,0 | 1,0 | 1,0 | 1,0 | 0,1 | 0,1 |
| | $Cu^{2+}$ | 1,0 | 1,0 | 1,1 | 1,0 | 1,0 | 1,0 | 1,0 | 0,1 | 0,1 |
| | $Fe^{2+}$ | 1,- | 1,- | 1,- | 1,- | 1,0 | 1,0 | 1,0 | 0,1 | 0,1 |
| | $Fe^{3+}$ | 1,- | 1,- | 1,0 | 1,- | 1,0 | 1,0 | 1,0 | 0,1 | 0,1 |
| | $Pb^{2+}$ | 1,0 | 1,0 | 1,0 | 1,0 | 1,0 | 1,0 | 1,0 | 0,0 | 0,0 |
| | $Cr^{3+}$ | 1,0 | 1,0 | 1,0 | 1,0 | 1,0 | 1,0 | 1,0 | 0,0 | 0,0 |

**Table 1** Number of Voronoi diagrams constructed by binary combinations of metal ions on potassium ferrocyanide loaded gels.

| Metal ion | | Primary reactant ions | | | | | | | | |
|---|---|---|---|---|---|---|---|---|---|---|
| | | $Mn^{2+}$ | $Co^{2+}$ | $Ag^{+}$ | $Ni^{2+}$ | $Cu^{2+}$ | $Fe^{2+}$ | $Fe^{3+}$ | $Pb^{2+}$ | $Cr^{3+}$ |
| Secondary reactant ions | $Mn^{2+}$ | 1,0 | 1,0 | 1, - | 1,0 | 1,0 | 1,0 | 1, - | 0,1 | 0,1 |
| | $Co^{2+}$ | 1,0 | 1,0 | 1,1 | 1,0 | 1,0 | 1,0 | 1,1 | 0,1 | 0,1 |
| | $Ag^{+}$ | 1, - | 1, - | 1,0 | 1,- | 1,0 | 1,1 | 1,0 | 0,1 | 0,1 |
| | $Ni^{2+}$ | 1,0 | 1,0 | 1,1 | 1,0 | 1,0 | 1,0 | 1,1 | 0,1 | 0,1 |
| | $Cu^{2+}$ | 1,0 | 1,0 | 1,1 | 1,0 | 1,0 | 1,0 | 1,1 | 0,1 | 0,1 |
| | $Fe^{2+}$ | 1,- | 1,- | 1, - | 1,- | 1,- | 1,0 | 1,1 | 0,1 | 0,1 |
| | $Fe^{3+}$ | 1,- | 1,- | 1,1 | 1,- | 1,- | 1,0 | 1,0 | 0,1 | 0,1 |
| | $Pb^{2+}$ | 1,0 | 1,0 | 1,0 | 1,0 | 1,0 | 1,0 | 1,0 | 0,0 | 0,0 |
| | $Cr^{3+}$ | 1,0 | 1,0 | 1,0 | 1,0 | 1,0 | 1,0 | 1,0 | 0,0 | 0,0 |

**Table 2** Number of Voronoi diagrams constructed by binary combinations of metal ions on potassium ferricyanide loaded gels.

**Key to Table 1 and Table 2**: [0,0] both reactant ions fail to form Voronoi diagrams, [1,0] the primary reactant ions form a Voronoi diagram but the secondary reactant ions are unreactive, [0,1] the primary reactant ions fail to construct a Voronoi diagram but the secondary reactant ions form a complete Voronoi diagram, [1,1] Both sets of reactant ions form Voronoi diagrams sequentially, [1,-] The primary set of reactants forms a Voronoi diagram but the secondary reactant ions form an incomplete or non-permanent Voronoi diagram.

**Figure 1a.**

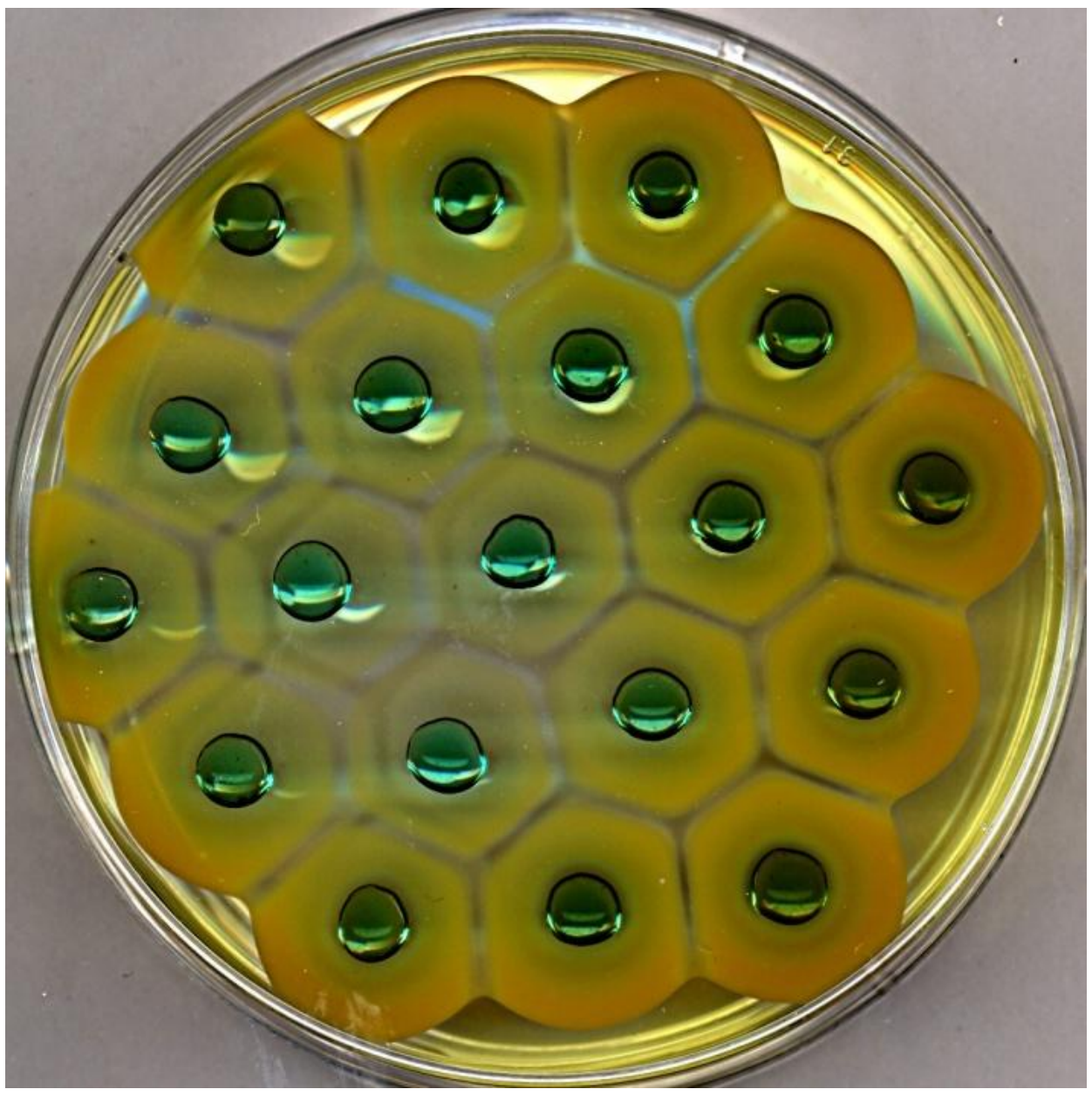

**Figure 1b.**

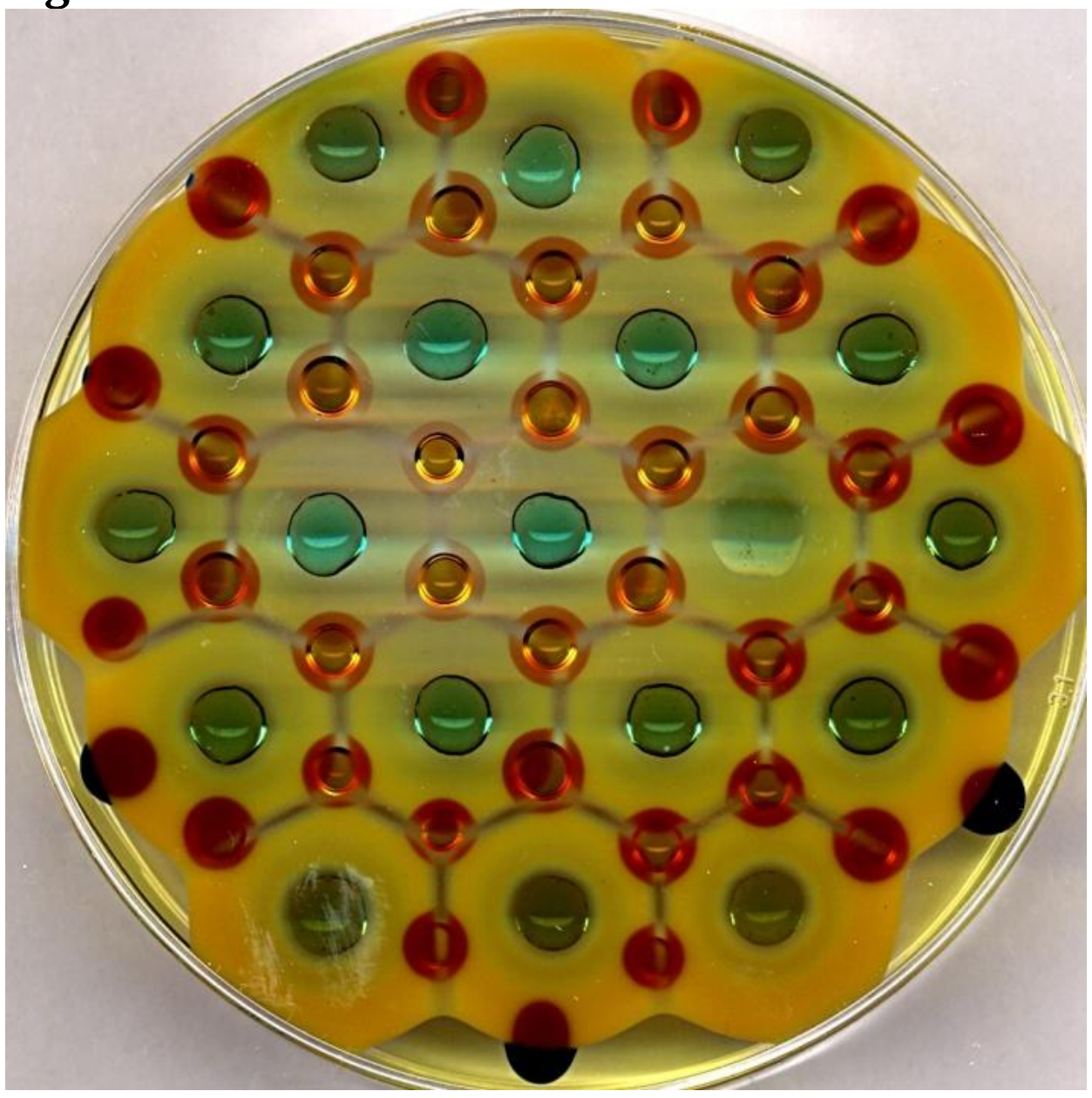

**Figure 1c.**

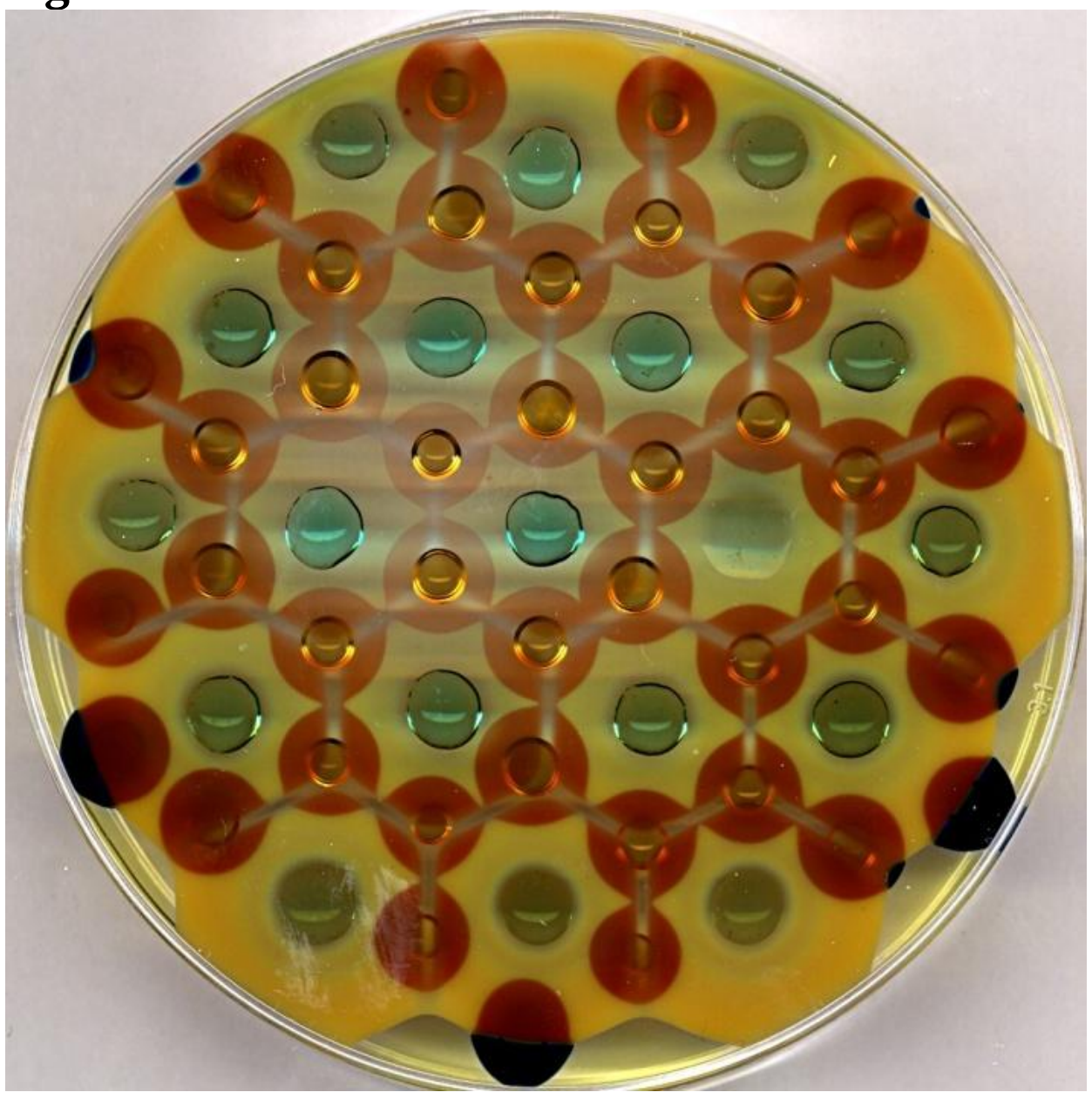

**Figure 1a.** A completed Voronoi diagram formed from the reaction of $Cu^{2+}$ ions on a potassium ferricyanide gel substrate. **b.** Addition of a secondary reagent in the form of $Fe^{2+}$ ions showing a change in colour of the original precipitate as the secondary reagent diffuses through the media. It is apparent that the colour is distinct from either of the original products if they were reacted separately on the gel **c.** In progress reaction showing the interaction of secondary reagent fronts without the formation of a permanent secondary Voronoi diagram. However, the intersection points of the overlapping fronts do constitute the formation of a secondary Voronoi diagram if extracted via further processing.

**Figure 2a**

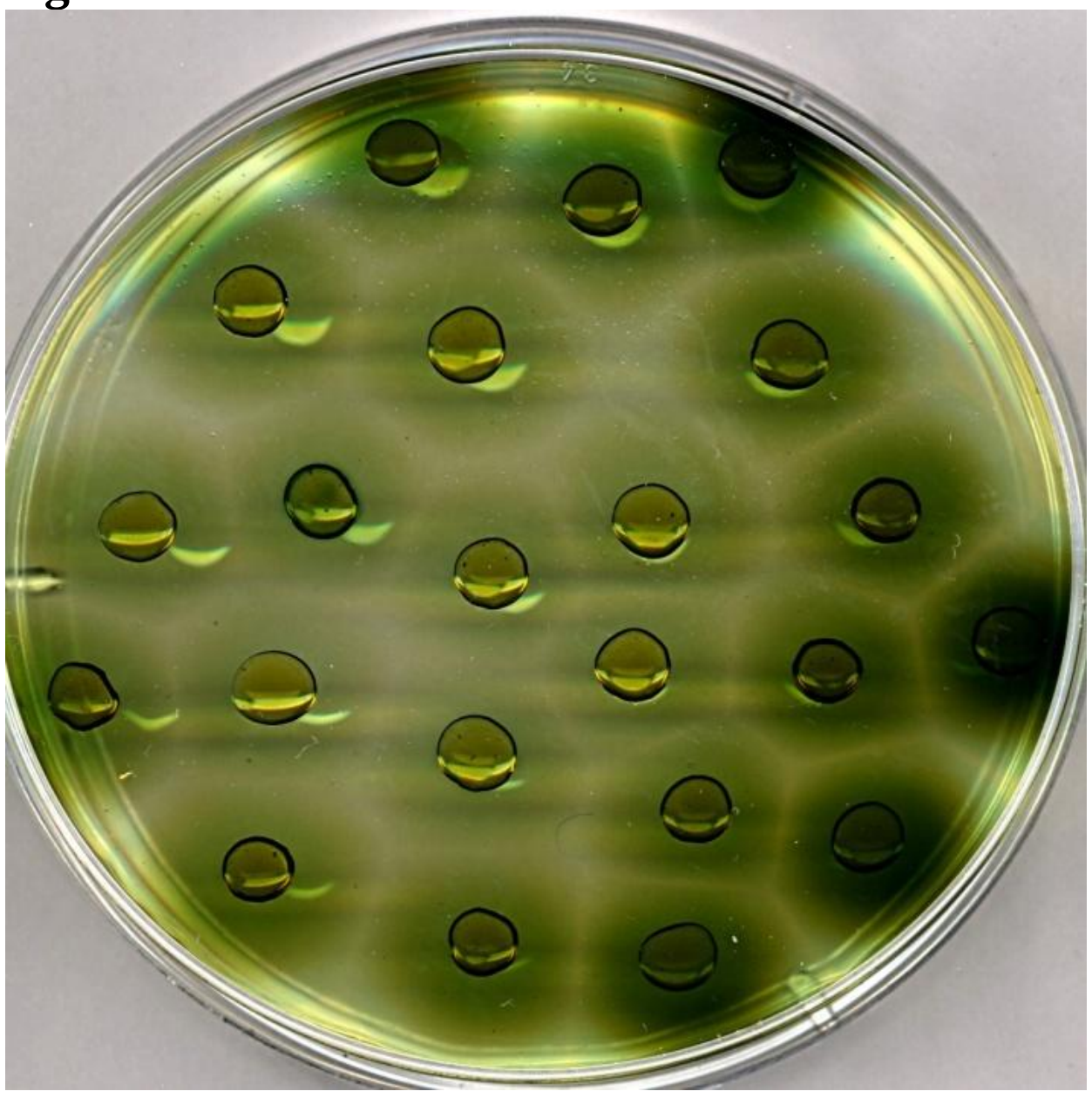

**Figure 2b**

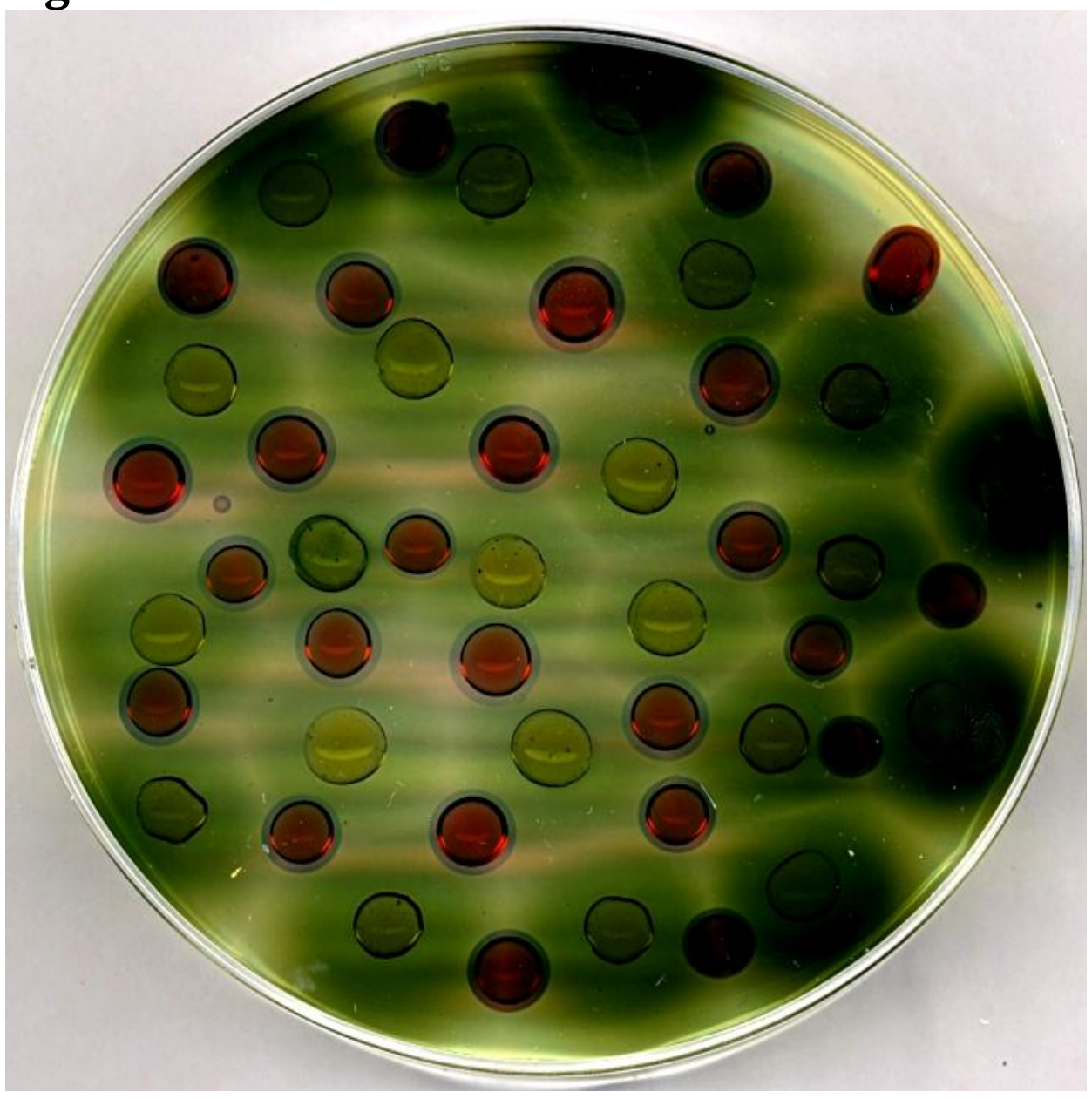

**Figure 2c**

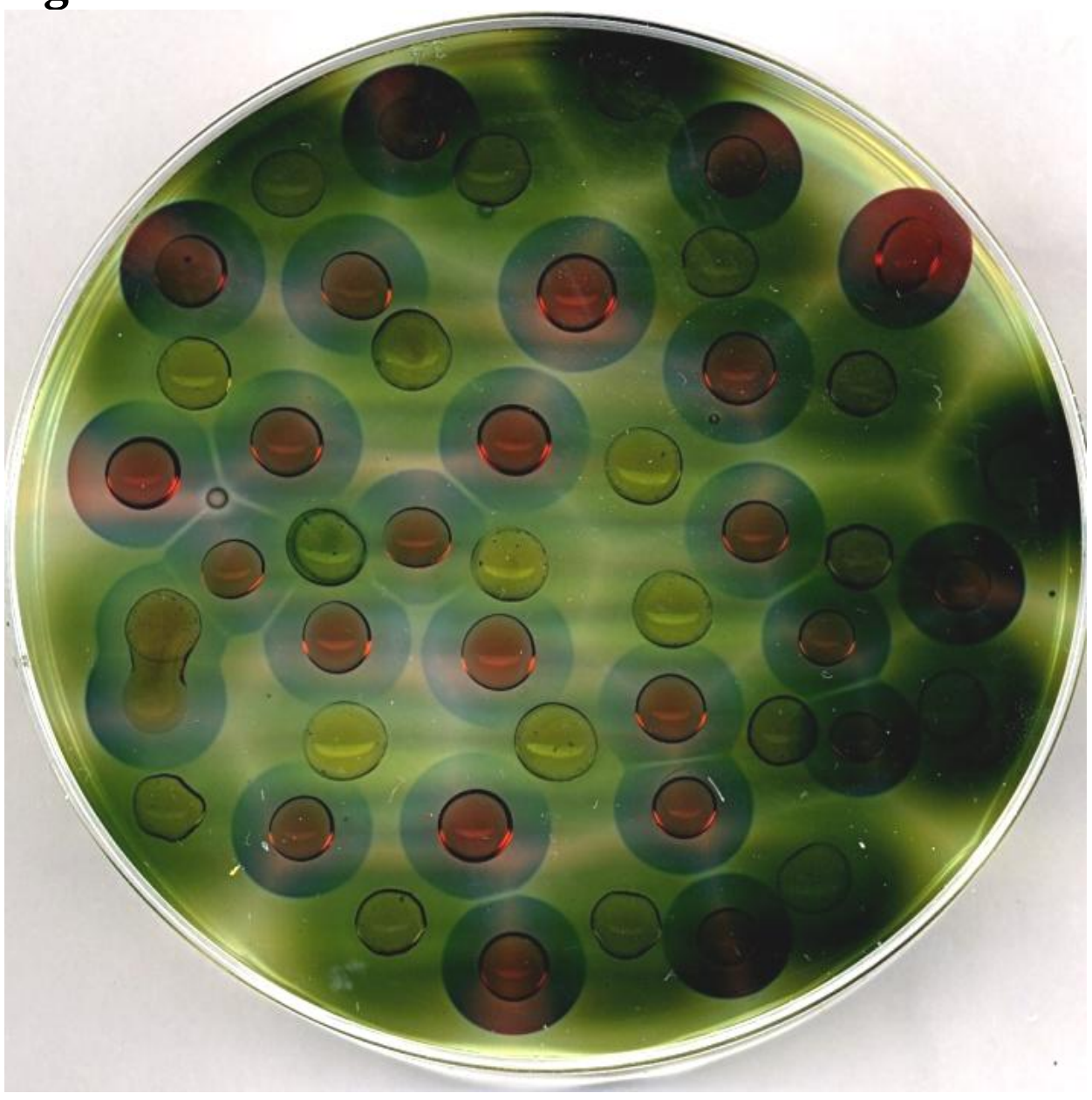

**Figure 2d**

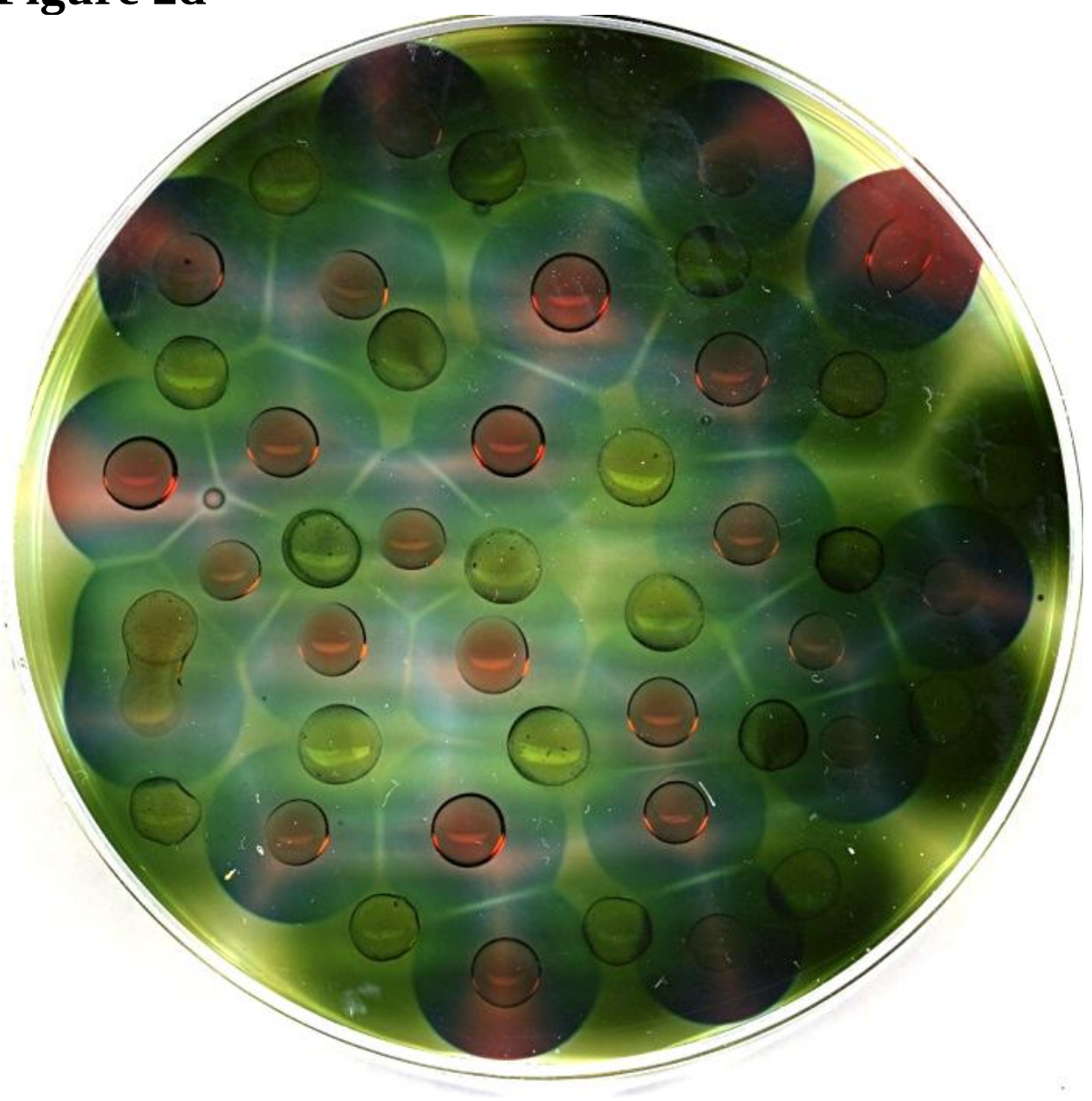

**Figure 2e**

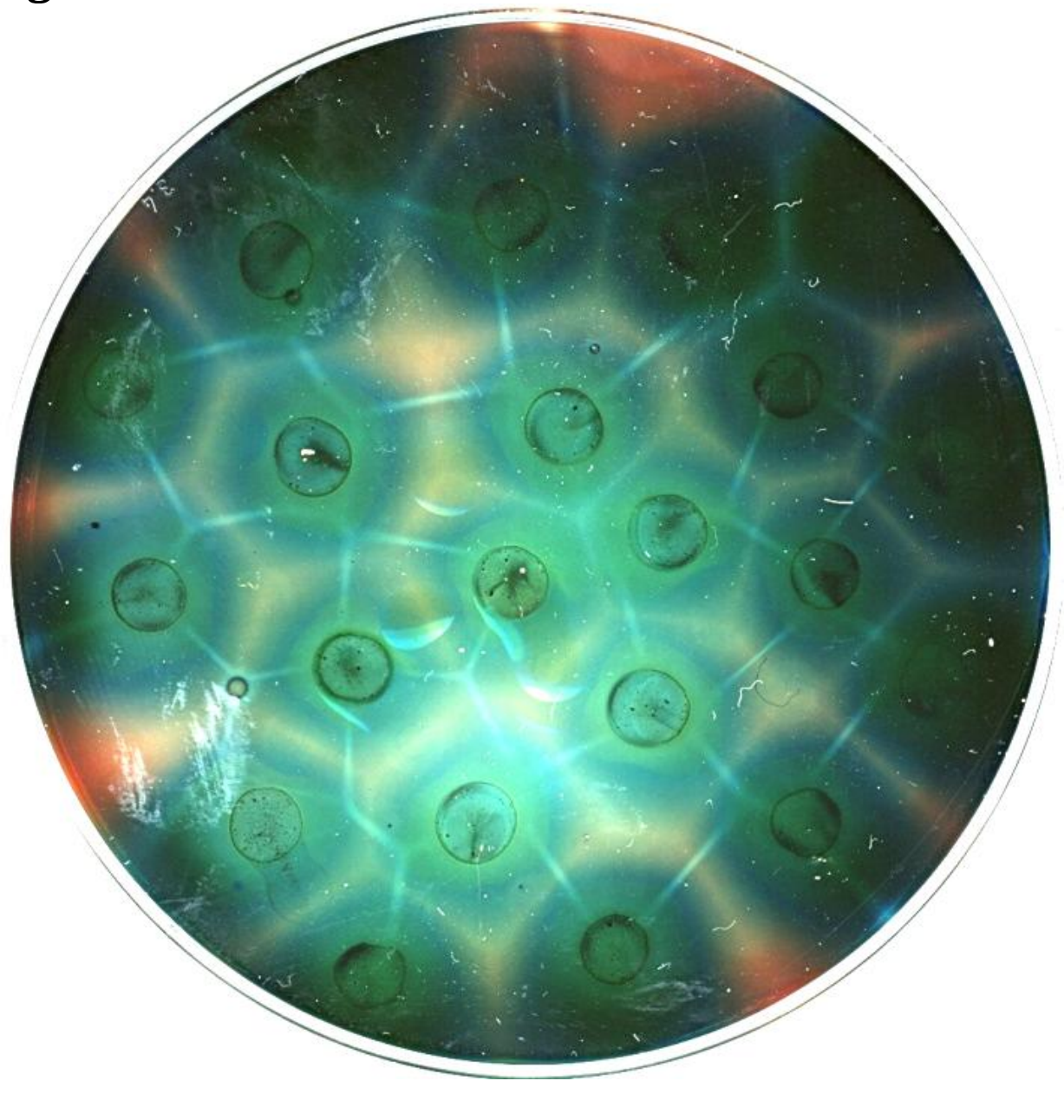

**Figure 2a.** Showing the completed Voronoi diagram where $Fe^{3+}$ ions are reacted on potassium ferricyanide gel substrate. **b.** The addition of $Co^{2+}$ ions to the gel substrate containing the original Voronoi diagram. **c.** In progress reaction showing the diffusion of $Co^{2+}$ ions and the secondary reaction with the substrate. **d.** Near complete reaction showing permanent Voronoi diagram bisector formation at points where the $Co^{2+}$ fronts interact. **e.** completed reaction showing the distinctive overlapping Voronoi diagram formation.

**Figure 3a**

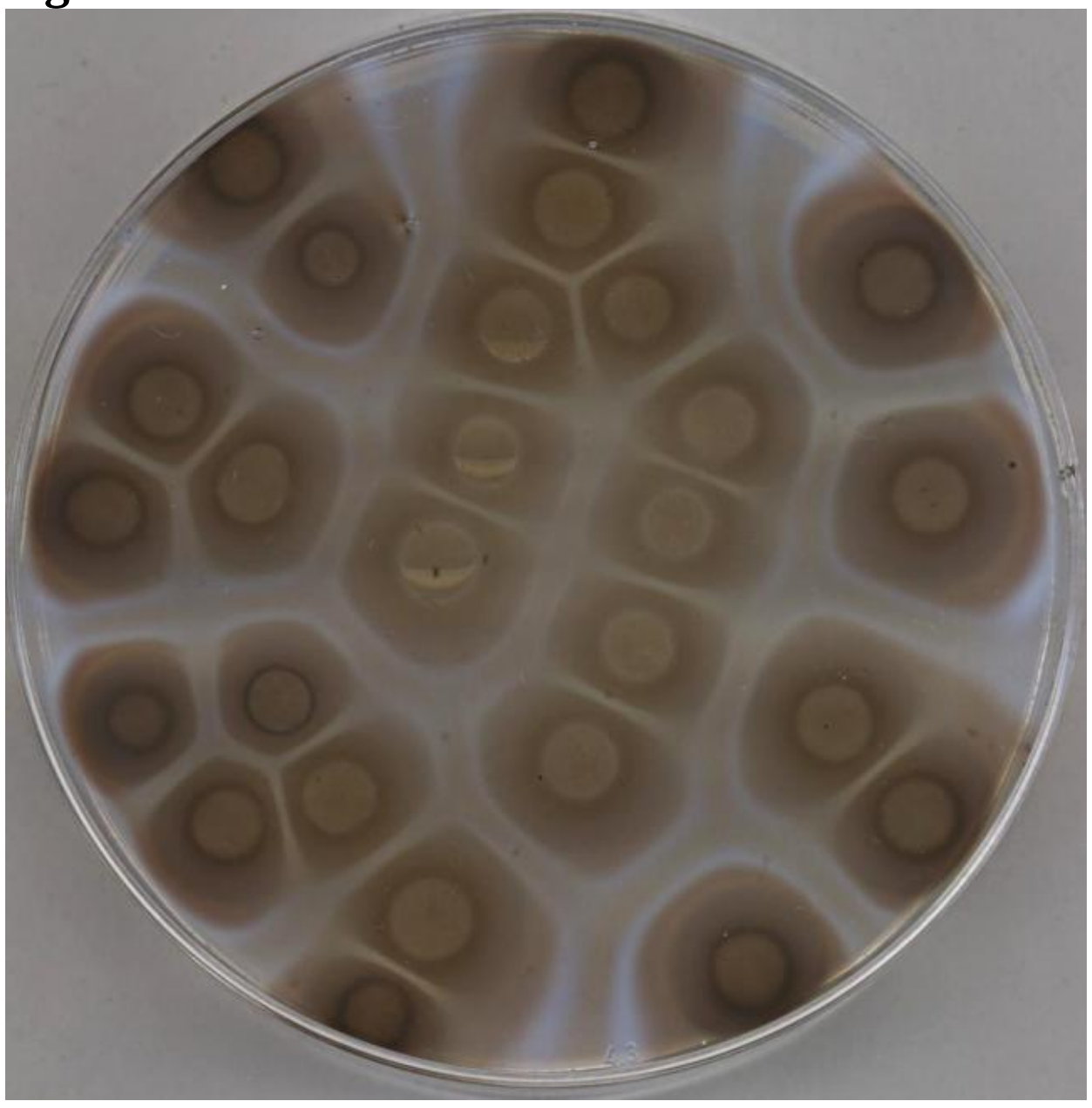

**Figure 3b**

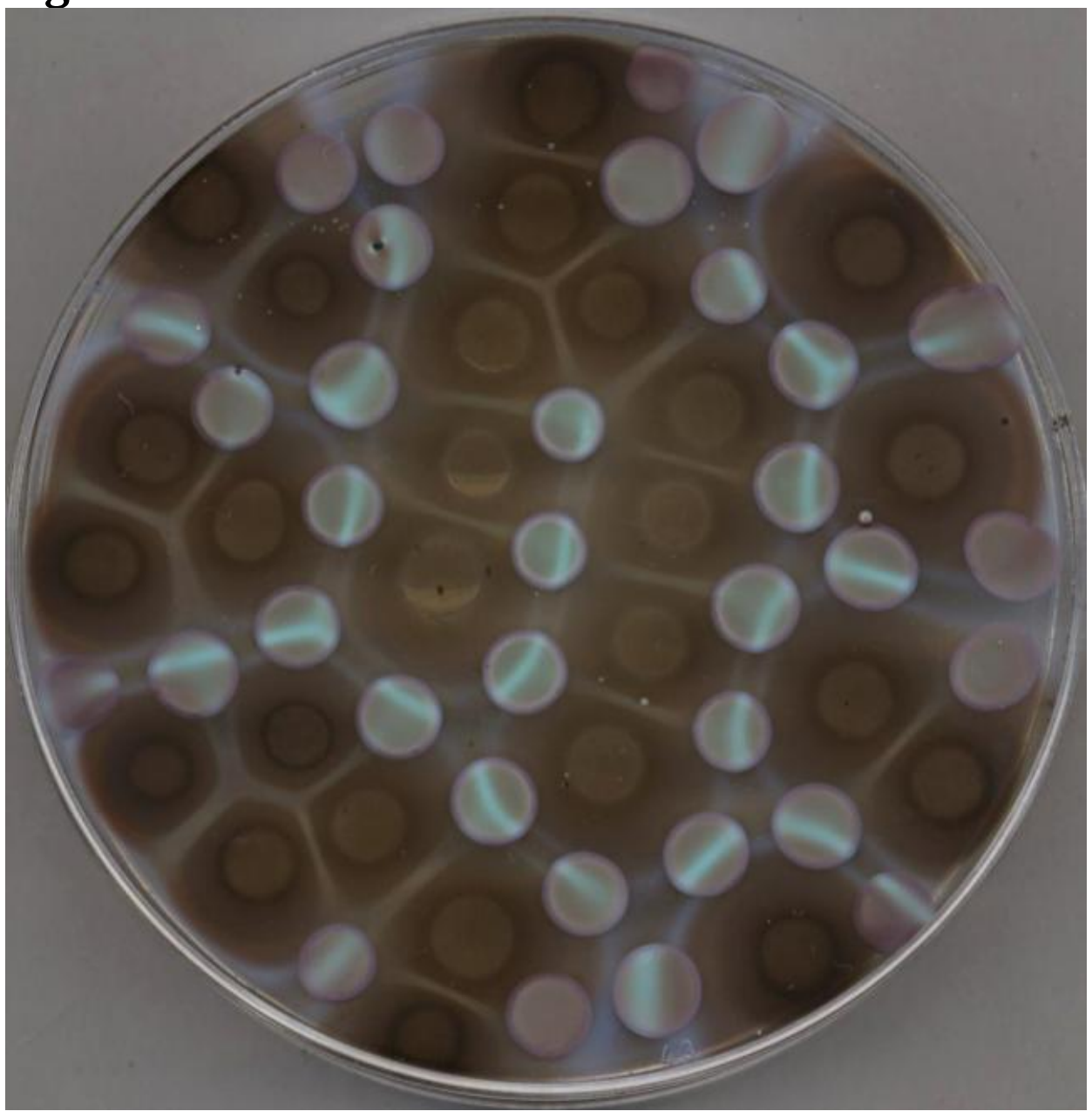

**Figure3c**

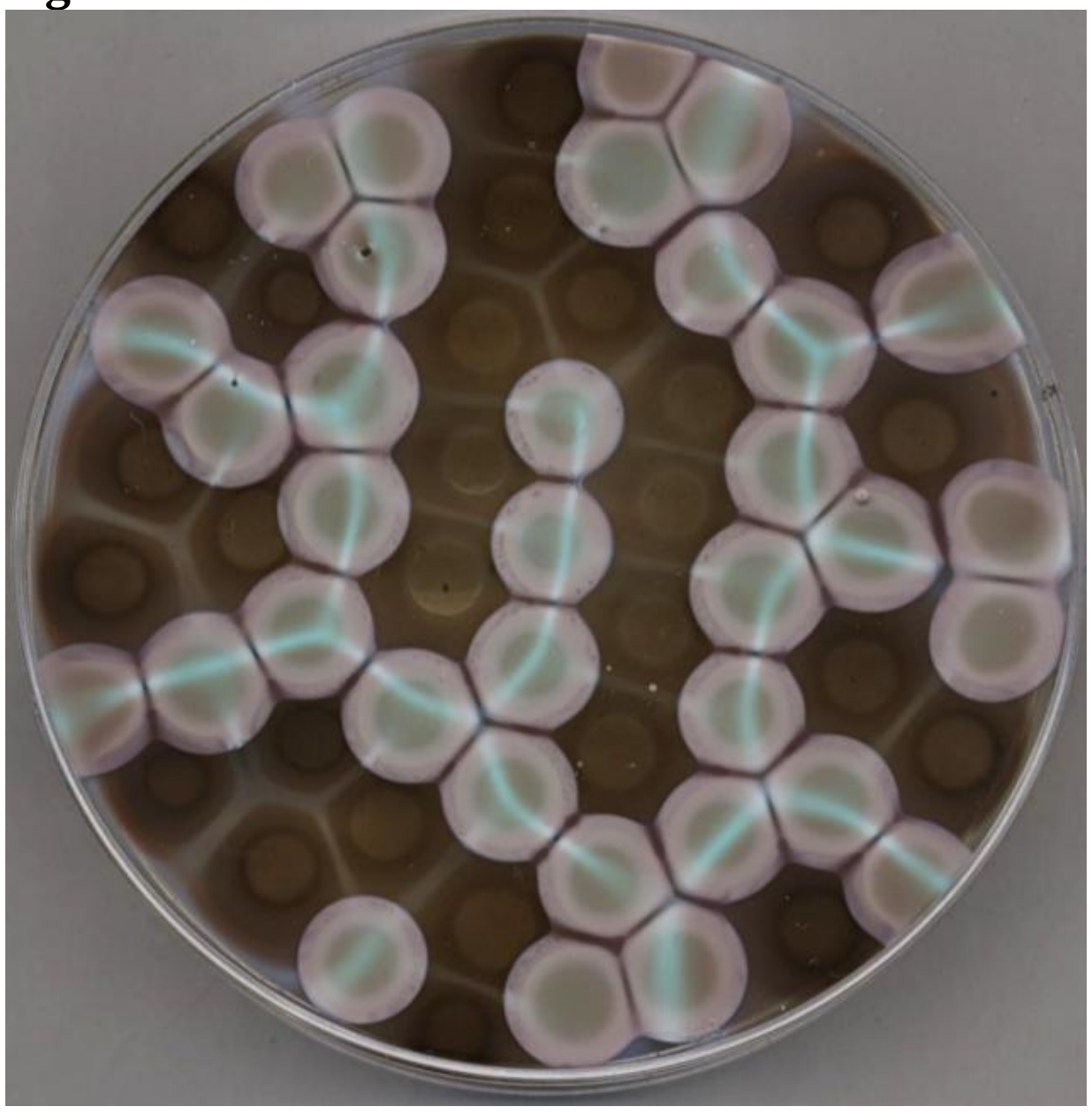

**Figure 3d**

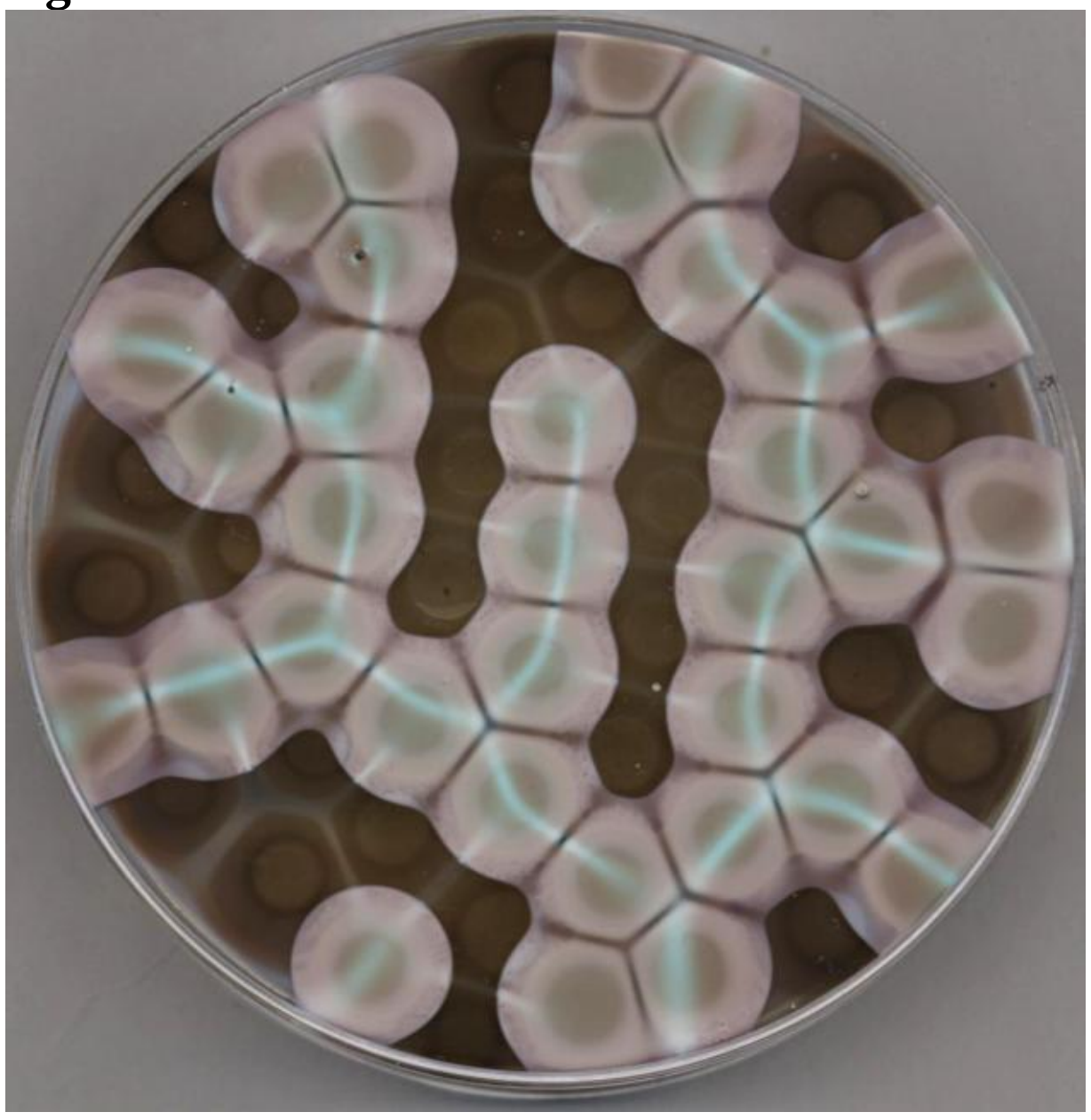

**Figure3e**

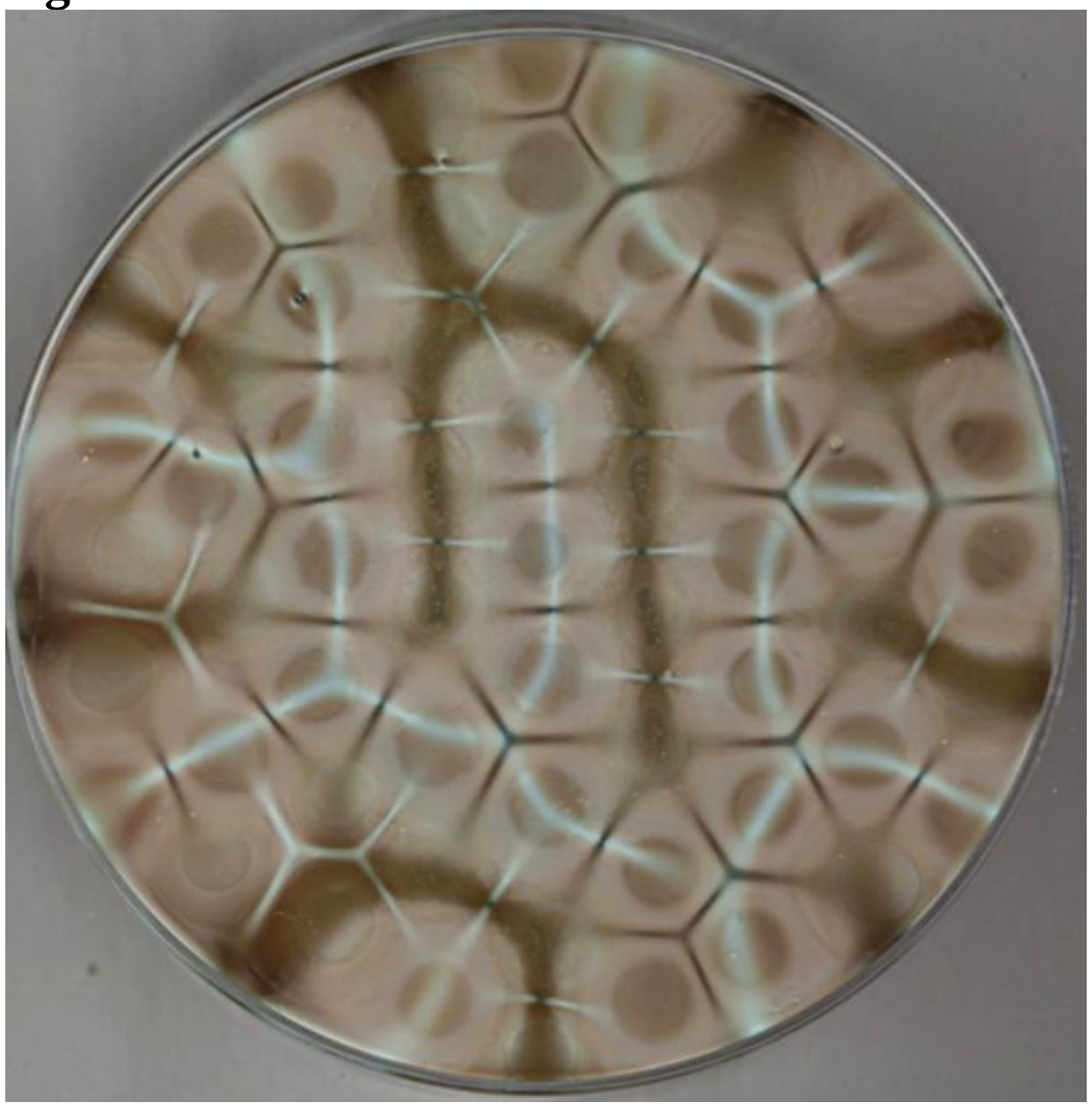

**Figure 3a**. Voronoi diagram formed by $Ag^{+}$ ions reacted on potassium ferrocyanide gel. **b.** Addition of Secondary $Co^{2+}$ reactant to the gel ubstrate containing the original Voronoi diagram. It can be seen that a physical precipitation reaction is intiated. **c.** 45 minutes into the reaction, $Co^{2+}$ diffusing fronts interact to form secondary Voronoi diagram bisectors (precipitate free regions). **d.** 1hour 30 minutes into the reaction all proximal sites have generated bisectors pertaining to a secondary Voronoi diagram. **e.** Final reaction showing formation of distinctive interacting Voronoi diagrams. The bisectors formed in these chemical reactions inherently code for distance. Therefore, points/sites separated by larger distances produce wide bisectors. This is useful when used in optimal path planning and minimal data reduction via the skeletonisation of planar shapes.

**Figure 4a**

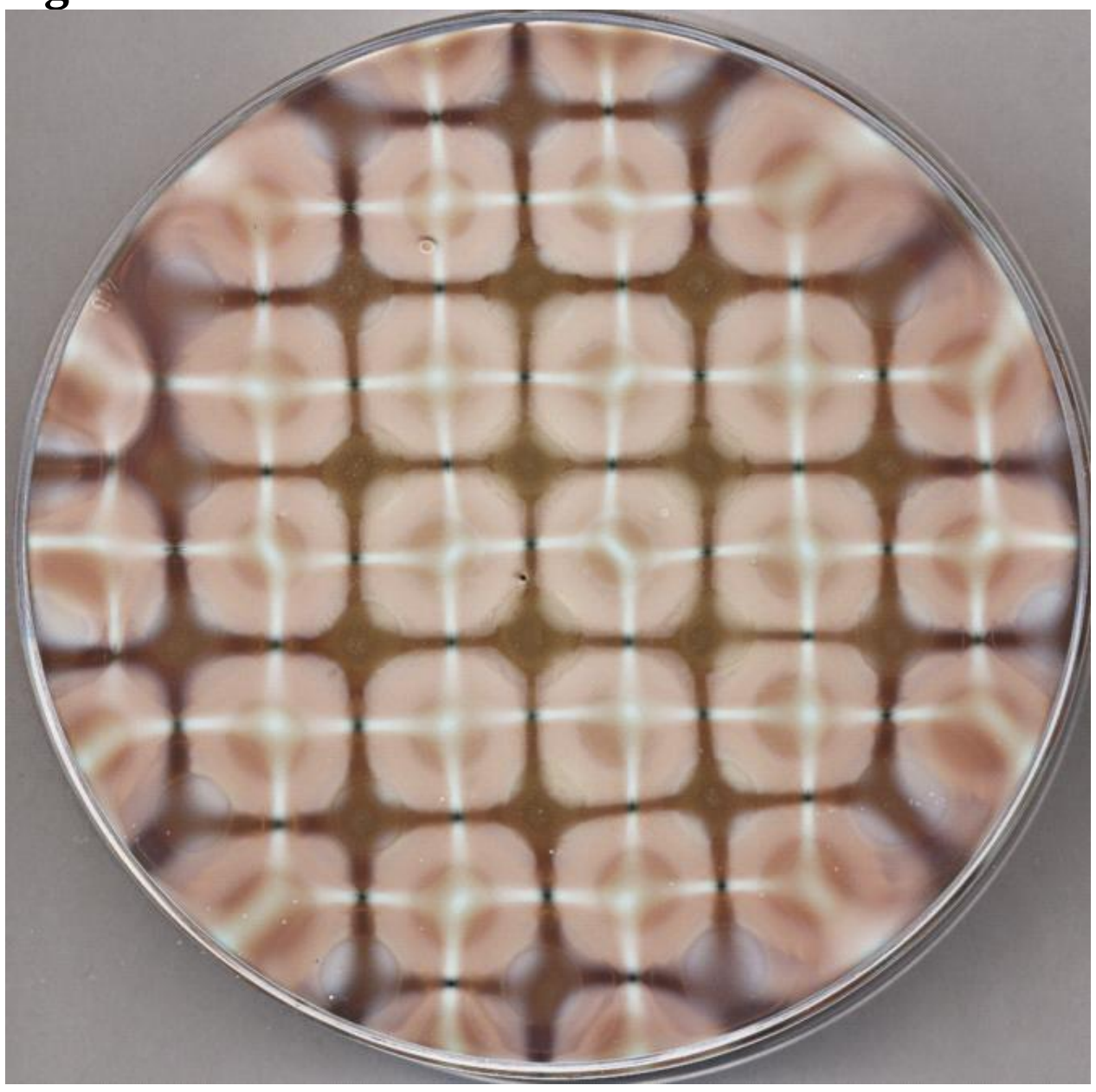

**Figure 4b**

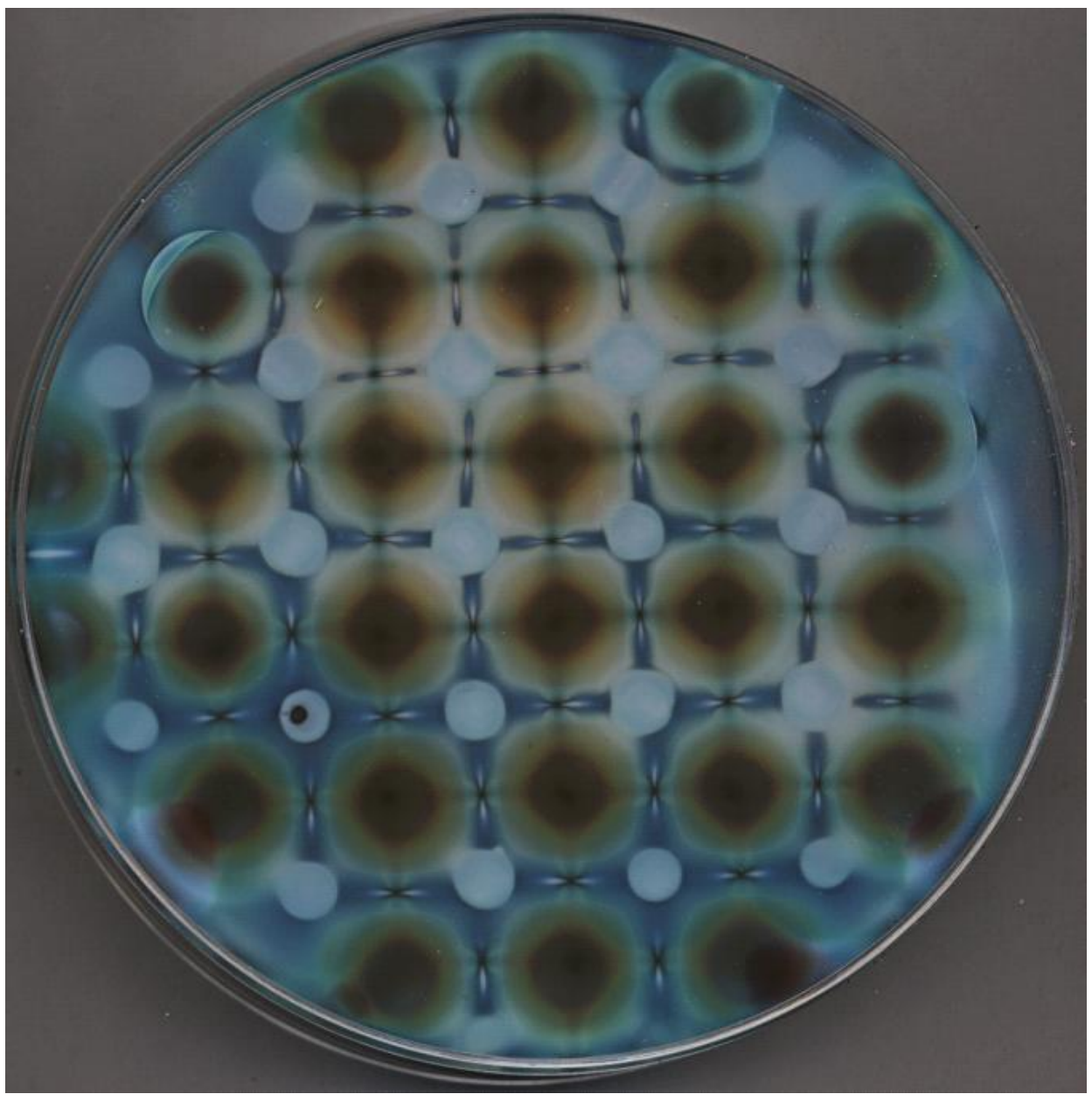

**Figure 4c**

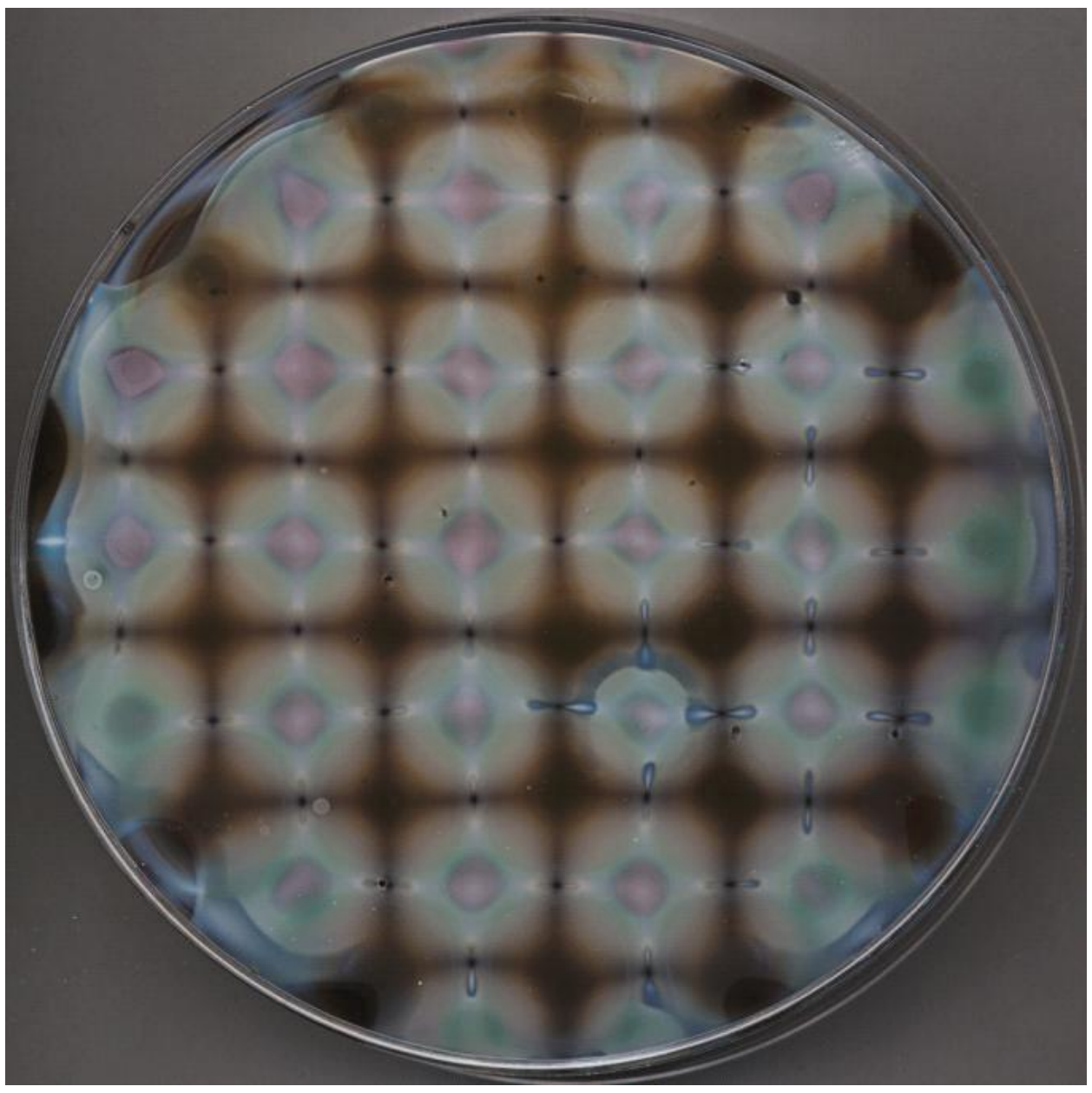

**Figure 4d**

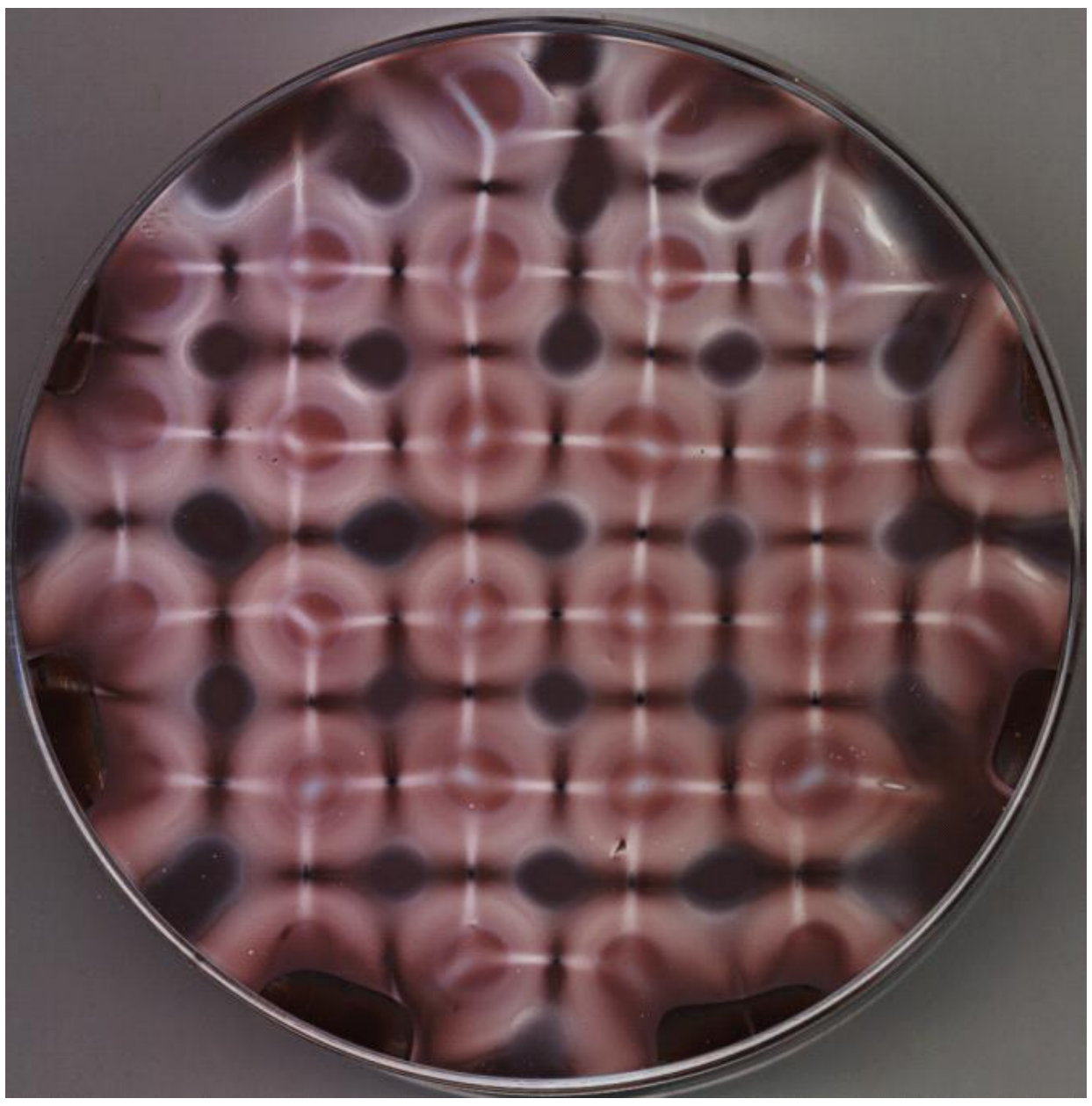

**Figure 4e**

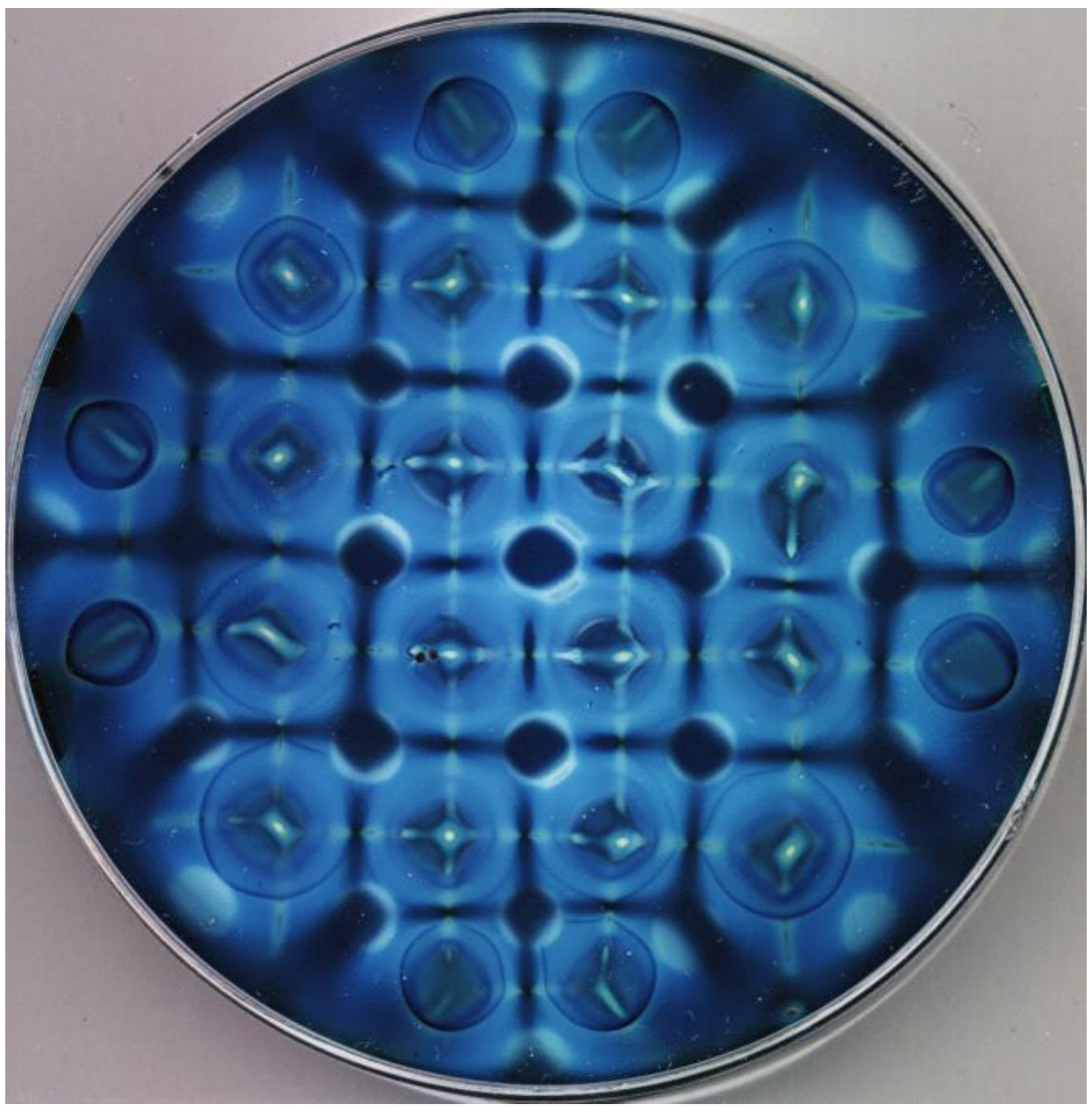

**Figure 4f**

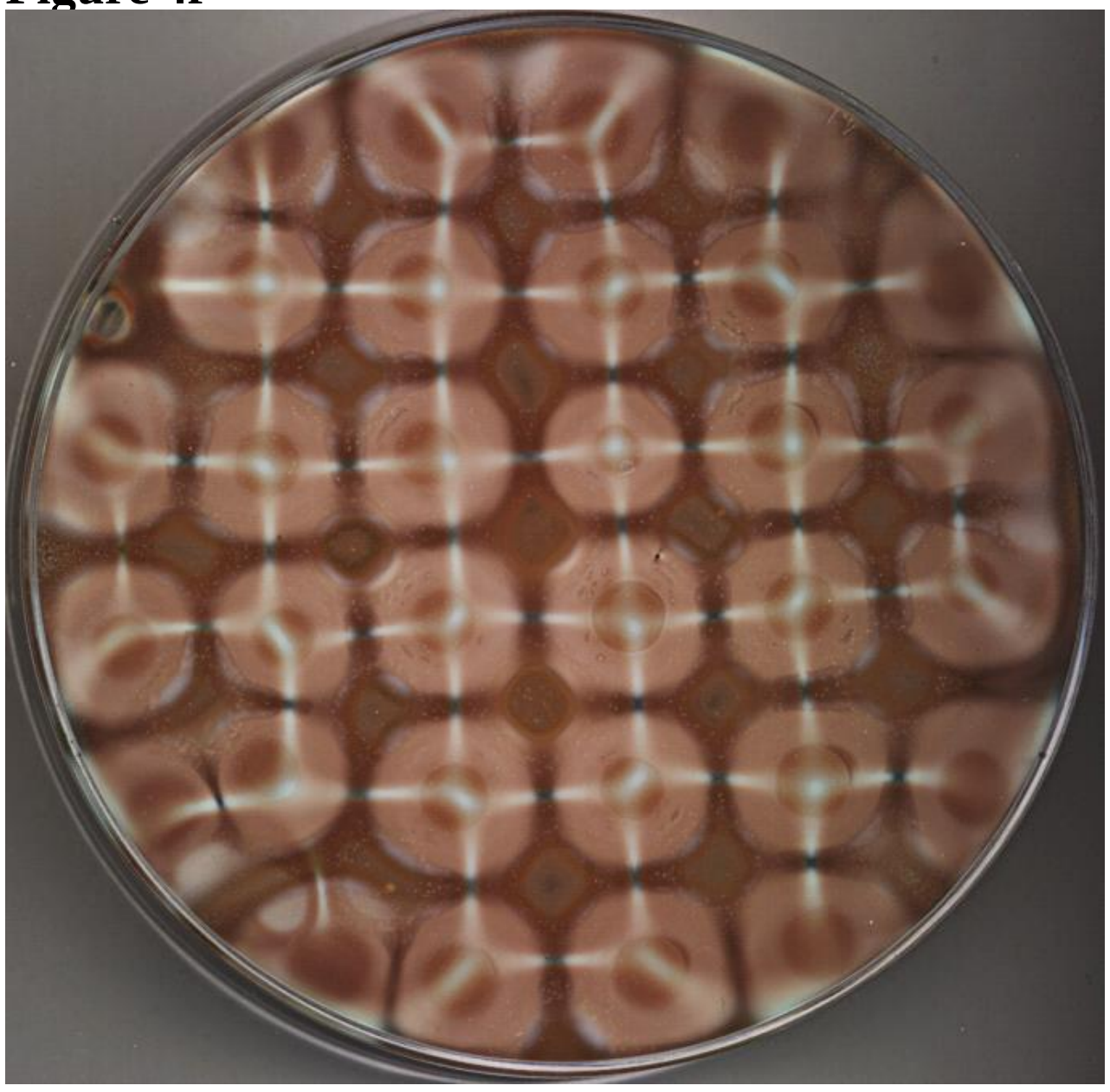

**Figure 4.** Completed sequential Voronoi diagram calculations for **a)** $Ag^{+}$ ($1^{st}$ reactant) $Cu^{2+}$ ($2^{nd}$ reactant) **b)** $Ag^{+}$ ($1^{st}$ reactant) and $Co^{2+}$ ($2^{nd}$ reactant) **c)** $Ag^{+}$ ($1^{st}$ reactant) and $Mn^{2+}$ ($2^{nd}$ reactant) reacted on potassium ferrocyanide gel. **d)** $Ag^{+}$ and $Co^{2+}$ **e)** $Ag^{+}$ and $Fe^{3+}$ and **f)** $Ag^{+}$ and $Cu^{2+}$ on potassium ferricyanide gel.

**Figure 5a**

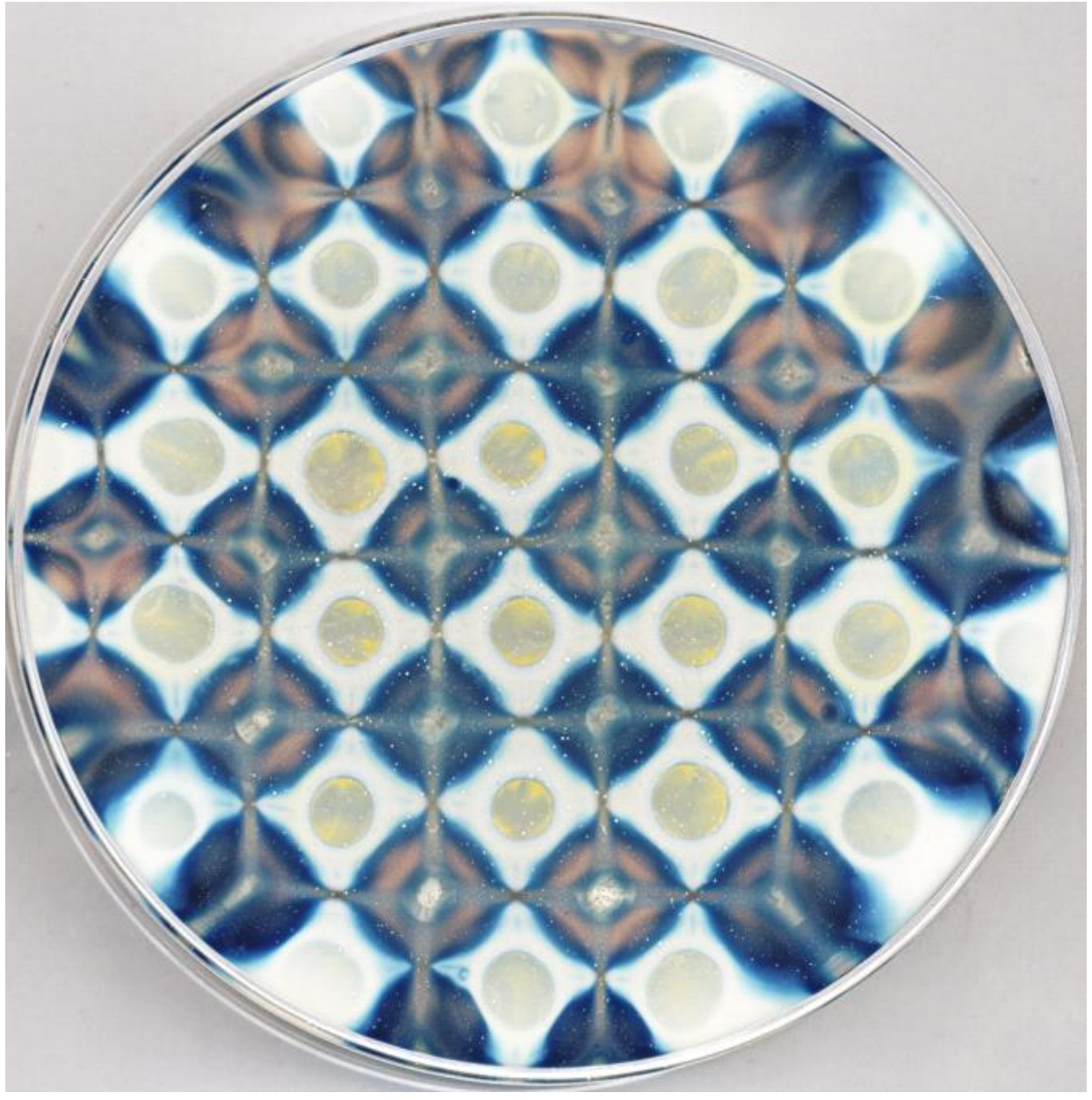

**Figure 5b**

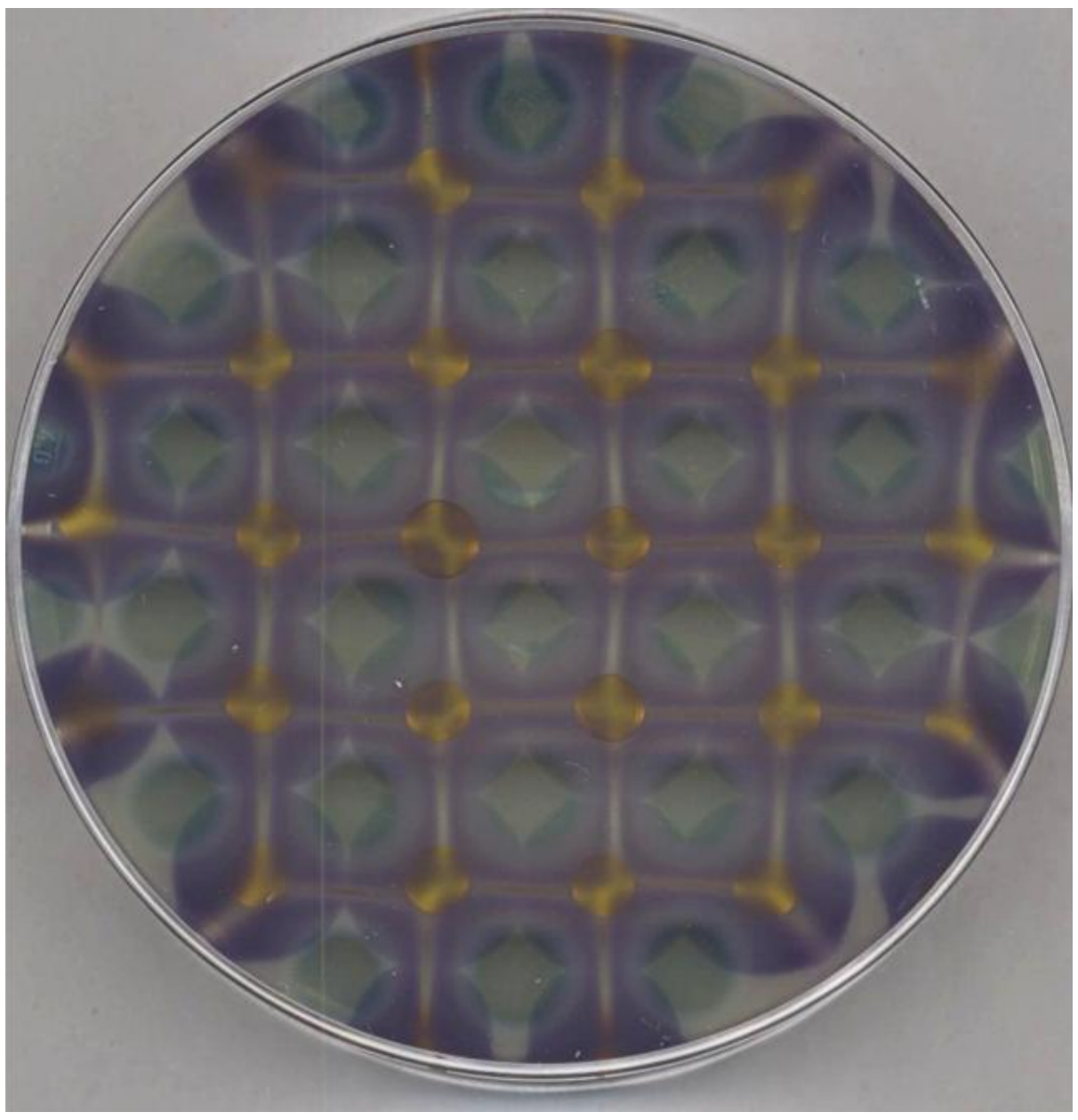

**Figure 5c**

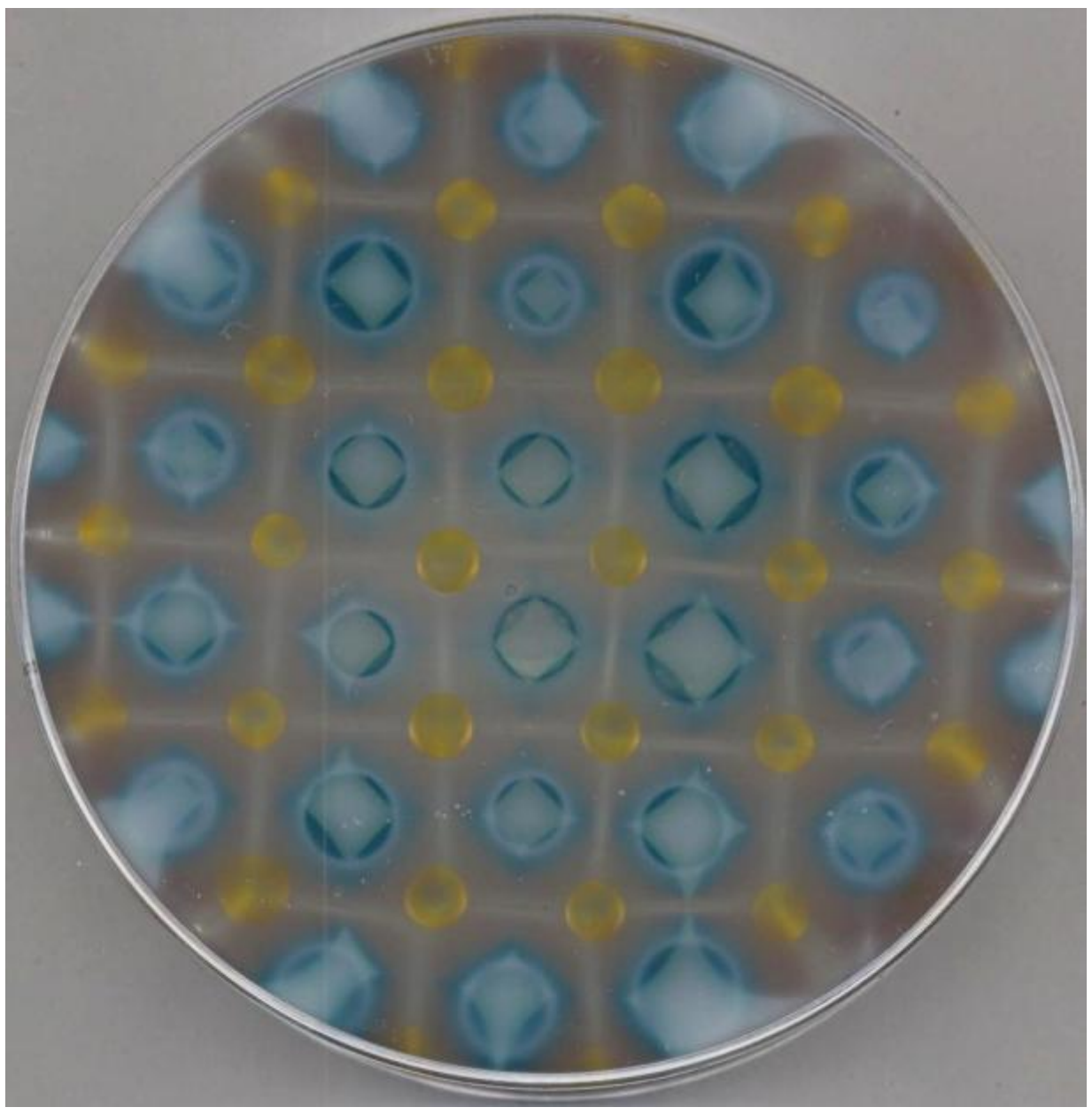

**Figure 5d**

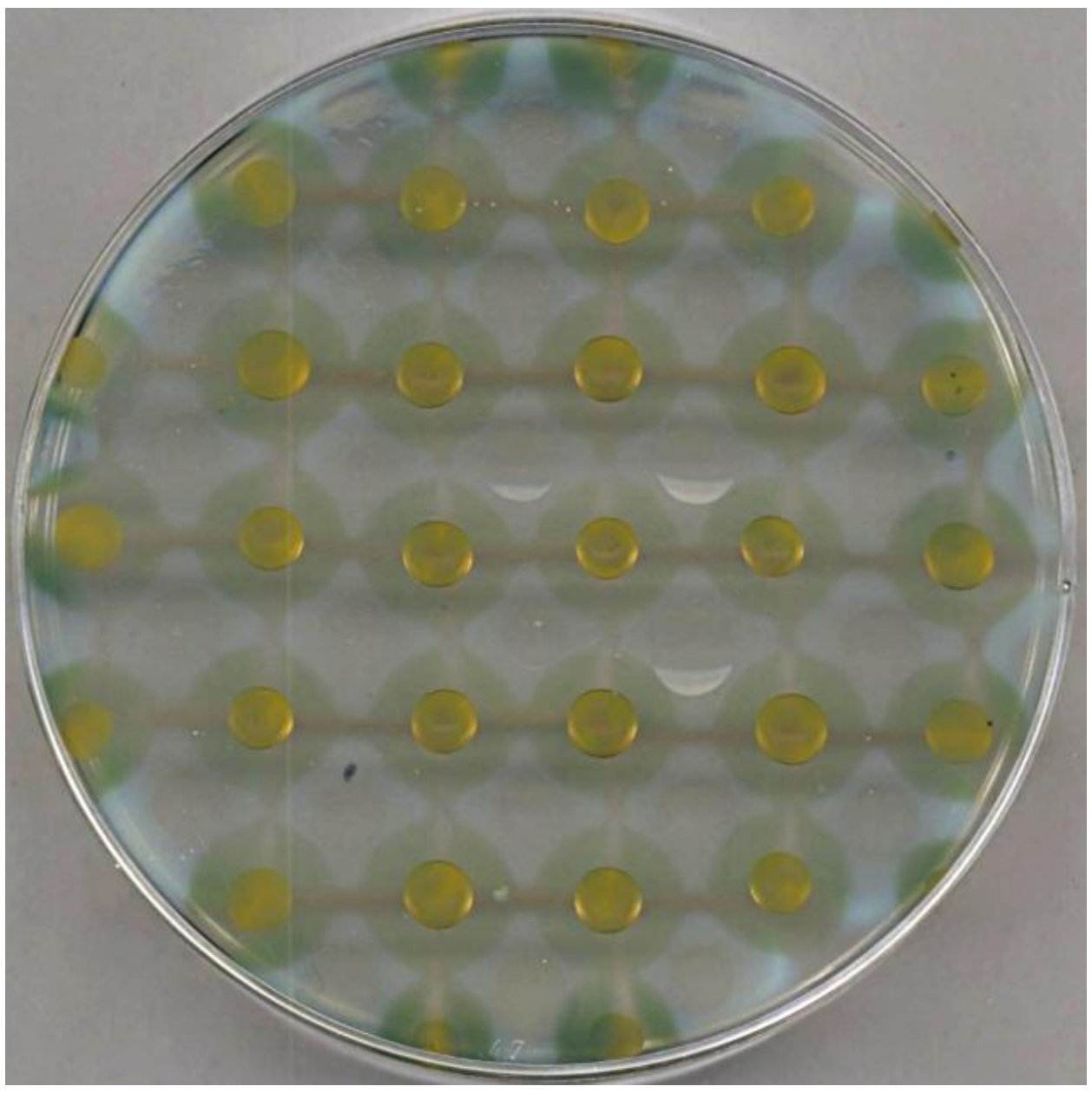

**Figure 5** Shows the completed reactions between **a)** $Co^{2+}$ ions (primary reactant) and $Fe^{2+}$ ions (secondary reactant), **b)** $Fe^{2+}$ ions (primary reactant) and $Ag^{+}$ ions (secondary reactant), **c)** $Mn^{2+}$ ions (primary reactant) and $Fe^{2+}$ ions (secondary reactant) and **d)** $Ni^{2+}$ ions (primary reactant) and $Fe^{2+}$ ions (secondary reactant) on a potassium ferrocyanide gel. These figures all show the incomplete formation of two interpenetrating Voronoi diagrams.

**Figure 6a**

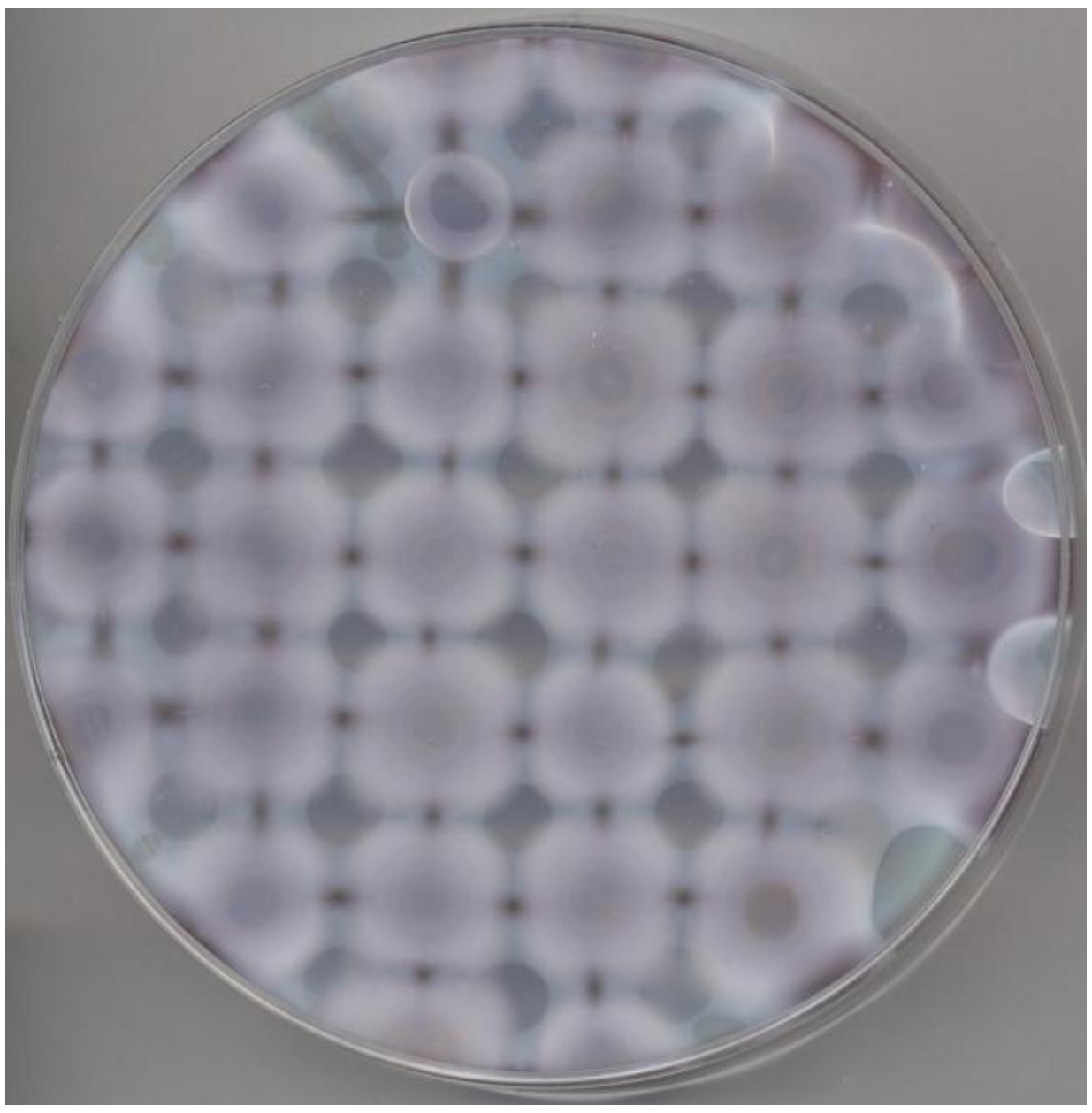

**Figure 6b**

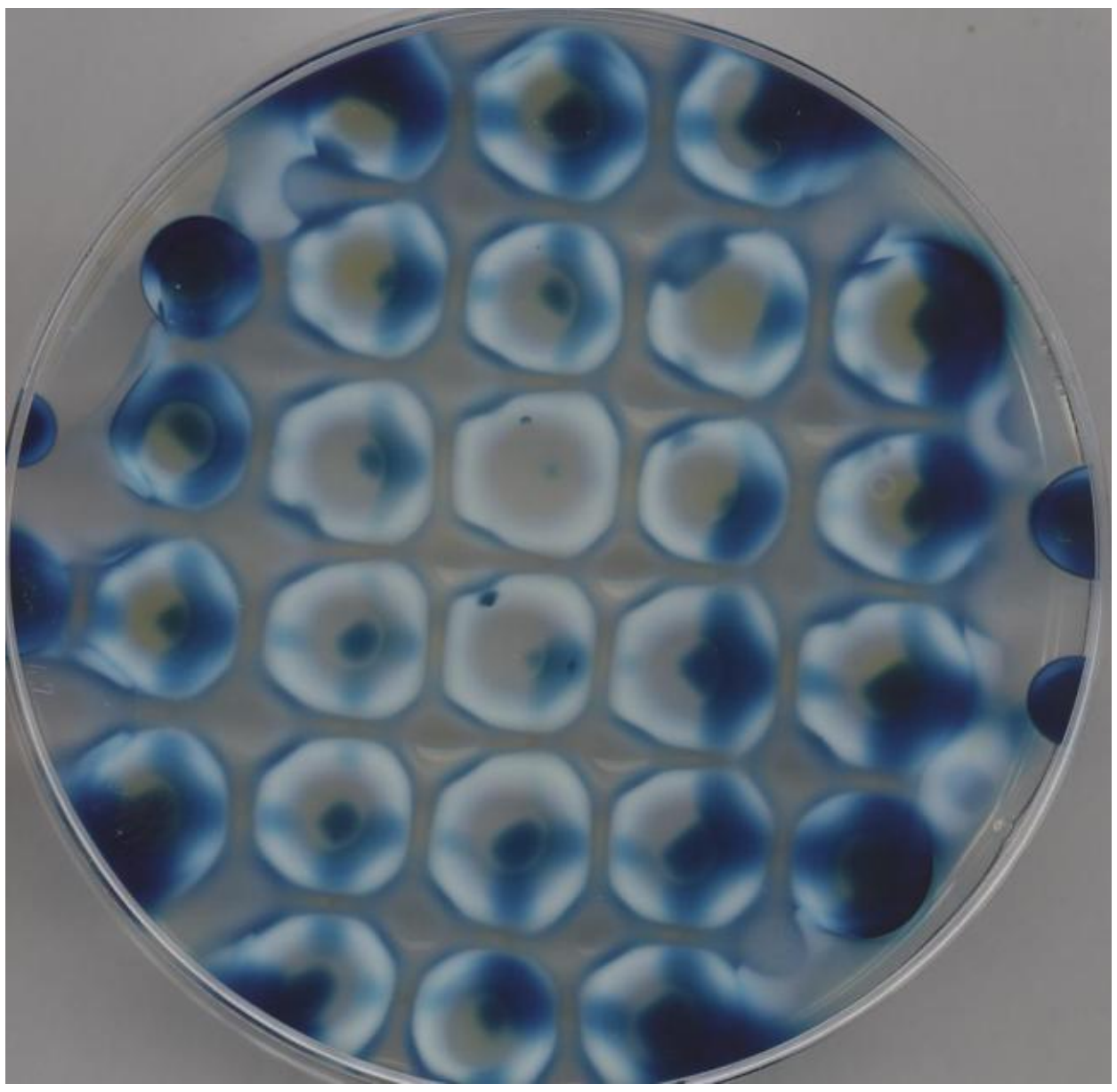

**Figure 6** Showing completed sequential Voronoi diagrams for **a)** the reaction of $Ni^{2+}$ and $Ag^{+}$ on potassium ferrocyanide gel and **b)** $Fe^{3+}$ reacted with $Ag^{+}$ on potassium ferrocyanide gel.